\documentclass[twocolumn]{aastex63}

\usepackage{amsmath}
\usepackage{natbib}
\usepackage{subfig}

\newcommand{\lowzr}{$1.3<z<1.5$}
\newcommand{\highzr}{$1.92<z<2.28$}
\newcommand{\lowz}{$z\sim1.4$}
\newcommand{\highz}{$z\sim2.1$}

\usepackage{lineno}
\DeclareCaptionFormat{cont}{#1#2#3\par}

\shorttitle{The Heavy Metal Survey: Elemental Abundances and Ages}
\shortauthors{Aliza G. Beverage}

\begin{document}

\title{The Heavy Metal Survey: The Evolution of Stellar Metallicities, Abundance Ratios, and Ages of Massive Quiescent Galaxies Since $z\sim2$}

\correspondingauthor{Aliza Beverage}
\email{abeverage@berkeley.edu}

\author[0000-0002-9861-4515]{Aliza G. Beverage}
\affiliation{Astronomy Department, University of California, Berkeley, CA 94720, USA}

\author[0000-0002-7613-9872]{Mariska Kriek}
\affiliation{Leiden Observatory, Leiden University, P.O. Box 9513, 2300 RA Leiden, The Netherlands}

\author[0000-0002-1714-1905]{Katherine A. Suess}
\altaffiliation{NHFP Hubble Fellow}
\affiliation{Kavli Institute for Particle Astrophysics and Cosmology and Department of Physics, Stanford University, Stanford, CA 94305, USA}

\author[0000-0002-1590-8551]{Charlie Conroy}
\affiliation{Center for Astrophysics \textbar\ Harvard \& Smithsonian, Cambridge, MA, 02138, USA}

\author[0000-0002-0108-4176]{Sedona H. Price}
\affiliation{Department of Physics \& Astronomy and PITT PACC, University of Pittsburgh, Pittsburgh, PA 15260, USA}

\author[0000-0001-6813-875X]{Guillermo Barro}
\affiliation{University of the Pacific, Stockton, CA 90340 USA}

\author[0000-0001-5063-8254]{Rachel Bezanson}
\affiliation{Department of Physics \& Astronomy and PITT PACC, University of Pittsburgh, Pittsburgh, PA 15260, USA}

\author[0000-0002-8871-3026]{Marijn Franx}
\affiliation{Leiden Observatory, Leiden University, P.O. Box 9513, 2300 RA Leiden, The Netherlands}

\author[0000-0002-5337-5856]{Brian Lorenz}
\affiliation{Astronomy Department, University of California, Berkeley, CA 94720, USA}

\author[0000-0002-0463-9528]{Yilun Ma}
\affiliation{Department of Astrophysical Sciences, Princeton University, Princeton, NJ 08544, USA}

\author[0000-0002-8530-9765]{Lamiya A. Mowla}
\affiliation{Astronomy Department, Whitin Observatory, Wellesley College, 106 Central Street, Wellesley, MA 02481, USA}

\author[0000-0002-7075-9931]{Imad Pasha}
\affiliation{Department of Astronomy, Yale University, New Haven, CT 06511, USA}

\author[0000-0002-8282-9888]{Pieter van Dokkum}
\affiliation{Department of Astronomy, Yale University, New Haven, CT 06511, USA}

\author[0000-0002-6442-6030]{Daniel R. Weisz}
\affiliation{Astronomy Department, University of California, Berkeley, CA 94720, USA}




\begin{abstract}
We present the elemental abundances and ages of 19 massive quiescent galaxies at $z\sim1.4$ and $z\sim2.1$ from the Keck Heavy Metal Survey. The ultra-deep LRIS and MOSFIRE spectra were modeled using a full-spectrum stellar population fitting code with variable abundance patterns. The galaxies have iron abundances between [Fe/H] = -0.5 and -0.1~dex, with typical values of $-0.2$ [$-0.3$] at $z\sim1.4$ [$z\sim2.1$]. We also find a tentative $\log\sigma_v$-[Fe/H] relation at $z\sim1.4$. The magnesium-to-iron ratios span [Mg/Fe]$\,=0.1$\,--\,$0.6$~dex, with typical values of $0.3$ [$0.5$]~dex at $z\sim1.4$ [$z\sim2.1$]. The ages imply formation redshifts of $z_{\rm form}=2-8$. Compared to quiescent galaxies at lower redshifts, we find [Fe/H] was $\sim0.2$~dex lower at $z=1.4-2.1$. We find no evolution in [Mg/Fe] out to $z\sim1.4$, though the $z\sim2.1$ galaxies are $0.2$~dex enhanced compared to $z=0-0.7$. A comparison of these results to a chemical evolution model indicates that galaxies at higher redshift form at progressively earlier epochs and over shorter star-formation timescales, with the $z\sim2.1$ galaxies forming the bulk of their stars over 150~Myr at $z_{\rm form}\sim4$. This evolution cannot be solely attributed to an increased number of quiescent galaxies at later times; several Heavy Metal galaxies have extreme chemical properties not found in massive galaxies at $z\sim0.0-0.7$. Thus, the chemical properties of individual galaxies must evolve over time. Minor mergers also cannot fully account for this evolution as they cannot increase [Fe/H], particularly in galaxy centers. Consequently, the build-up of massive quiescent galaxies since $z\sim2.1$ may require further mechanisms such as major mergers and/or central star formation.

\end{abstract}


\section{Introduction}
\label{sec:intro}

In the present-day universe, nearly all massive galaxies have quiescent stellar populations. Archaeological studies have revealed that they formed the bulk of their stars rapidly at $z>2$ \citep[e.g.,][]{gallazzi_ages_2005, thomas_epochs_2005, mcdermid_atlas3d_2015}. Thus, these studies imply that quiescent galaxies should already exist when the universe was only a fraction of its current age. 

Quiescent galaxies have indeed been identified and studied out to $z\sim5$ \citep[e.g.,][]{franx_significant_2003, daddi_passively_2005, cimatti_old_2004, kriek_spectroscopic_2006, cimatti_gmass_2008, kriek_ultra-deep_2009, van_de_sande_stellar_2013, belli_velocity_2014, glazebrook_massive_2017, schreiber_near_2018, valentino_quiescent_2020, nanayakkara_population_2023, carnall_massive_2023, antwi-danso_feniks_2023} and are thought to dominate the massive galaxy population out to $z=2.5$ \citep[e.g.][]{muzzin_evolution_2013,tomczak_galaxy_2014,mcleod_evolution_2021}. Interestingly, quiescent galaxies at higher redshifts are significantly more compact than the population at $z\sim0$, with their half-mass radii growing by a factor of three since $z\sim2$ \citep{daddi_passively_2005, dokkum_confirmation_2008, damjanov_red_2009, van_der_wel_3d-hstcandels_2014, mowla_cosmos-dash_2019, suess_half-mass_2019, suess_half-mass_2019-1,miller_color_2023}. Distant quiescent galaxies, however, have central mass densities similar to their nearby analogs, and the size difference is primarily attributed to the larger envelopes of the lower-redshift galaxies \citep[e.g.,][]{bezanson_relation_2009,hopkins_compact_2009, van_dokkum_growth_2010, barro_structural_2017, van_de_sande_stellar_2013, ji_reconstructing_2022}. Furthermore, the color gradients of quiescent galaxies become stronger with time, with galaxy outskirts becoming progressively bluer than the centers \citep{suess_half-mass_2019, suess_half-mass_2019-1, suess_color_2020, miller_color_2023}.

One way to understand the structural evolution of massive quiescent galaxies is through the ``inside-out'' growth scenario, in which compact galaxies still grow via gas-poor minor mergers after they have become quiescent \citep[e.g.,][]{bezanson_relation_2009, naab_minor_2009, van_dokkum_growth_2010, oser_cosmological_2012}. An alternative explanation is suggested by the fact that the stellar mass function shows a tenfold increase in the number of massive quiescent galaxies since $z\sim2$ \citep{mcleod_evolution_2021}. If galaxies that quench at lower redshift have larger sizes and stronger color gradients, the observed evolution of \textit{average} properties could instead be explained by population growth \citep[i.e., progenitor bias; e.g.,][]{khochfar_dry_2009, van_dokkum_growth_2010, carollo_newly_2013, poggianti_evolution_2013}. The relative importance of minor mergers vs progenitor bias remains a central question in massive galaxy evolution. 

The evolution in elemental abundances and stellar population properties over cosmic time provides a unique insight into the assembly and star-formation histories of massive quiescent galaxies \citep[e.g.,][]{matteucci_abundance_1994,trager_stellar_2000, conroy_early-type_2014, choi_assembly_2014, maiolino_re_2019,beverage_elemental_2021, peng_strangulation_2015, spitoni_new_2017, trussler_both_2020}. Unfortunately, measuring elemental abundances beyond the low-redshift universe poses a considerable challenge. At higher redshifts, key absorption features are faint and shifted to near infrared (NIR) wavelengths. Recently, ultra-deep spectroscopic surveys have started pushing these measurements beyond the low-redshift universe and find that the chemical properties of massive quiescent galaxies at $z\sim0.7$ closely resemble today's massive early-type galaxies \citep{choi_assembly_2014, leethochawalit_evolution_2018, beverage_carbon_2023, bevacqua_elemental_2023}. These studies show that the most massive galaxies were already in place six billion years ago, while the low-mass quiescent population have continued to grow since $z\sim0.7$ \citep[see ][]{leethochawalit_evolution_2019,beverage_carbon_2023}.
 
While minimal evolution is found out to $z\sim0.7$, the picture at higher redshift is less clear. The few existing stellar metallicity measurements of massive quiescent galaxies at $z\gtrsim1.4$ show systematically lower [Fe/H] compared to $z\sim0$, but there is large scatter depending on the methods used to derive the metallicities \citep{lonoce_old_2015, onodera_ages_2015,kriek_massive_2016, kriek_stellar_2019,carnall_stellar_2022, saracco_star_2023, zhuang_glimpse_2023}. At $z>2$, the only \textit{two} existing measurements show strong $\alpha$-enhancement, with [Mg/Fe]$\sim0.5-0.6$\,dex, indicative of extremely short star-formation timescales of $<100$\,Myr \citep{kriek_massive_2016, jafariyazani_resolved_2020}. To further clarify the picture at $z>1$, and assess the assembly of the quiescent galaxy population over the past 11 billion years, larger galaxy samples are needed.

To that end, we have executed the Keck Heavy Metal survey, an ultra-deep spectroscopic survey with LRIS \citep{oke_keck_1995, rockosi_low-resolution_2010} and MOSFIRE \citep{mclean_design_2010, mclean_mosfire_2012} on the Keck I Telescope \citep[see][]{kriek_heavy_2023}. This survey collected high-S/N spectra of 21 massive quiescent galaxies at $1.4\lesssim z\lesssim2.2$, covering multiple Balmer and metal absorption lines. 

The Heavy Metal survey has more than tripled the number of individual galaxies at $z\gtrsim1.4$ with measurements of stellar metallicities, ages, and elemental abundance ratios. Preliminary results of five Heavy Metal galaxies were presented in \citet{kriek_stellar_2019} and are updated in the current work. The survey design is presented in the companion survey paper \citep{kriek_heavy_2023}. In section 2 we describe our full-spectrum modeling technique. In section 3 we present the stellar population results. In section 4 we assess how the metal content of quiescent galaxies evolves between $z\sim2.2$\ and $z\sim0$ by comparing our results with the LEGA-C ($z\sim0.7$) and SDSS ($z\sim0$) surveys. In section 5 we consider the implications of our results in the build-up of the massive quiescent galaxy population. In section 6 we summarize our results. Throughout this work we assume a flat $\Lambda$CDM cosmology with $\Omega_{\rm m}= 0.3$ and $H_{\rm 0} = 70$\;km\,s$^{-1}$\,Mpc$^{-1}$.

\begin{figure*}
    \centering
    \includegraphics[width=\textwidth]{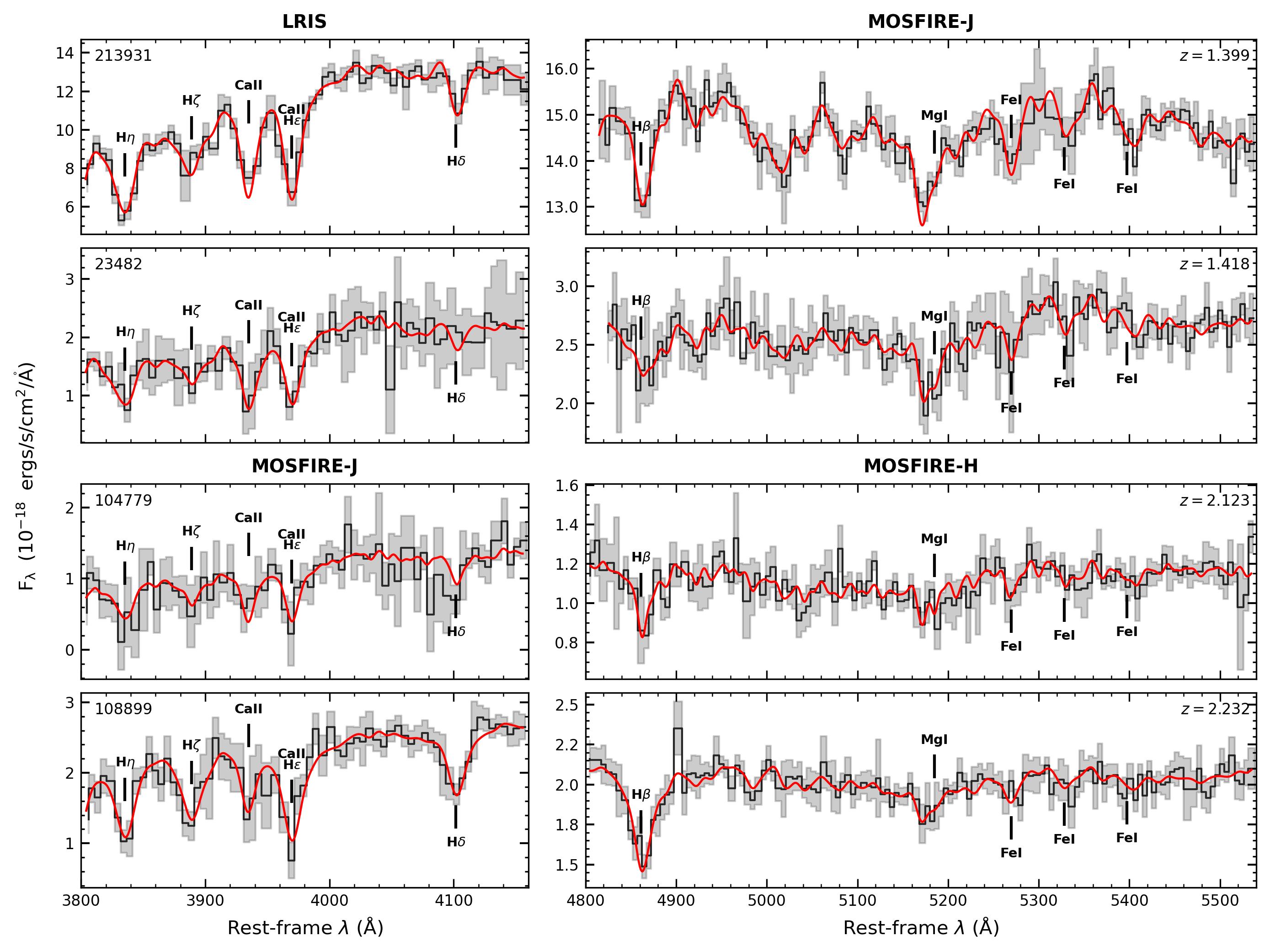}
    \caption{Spectra (black) and corresponding 1$\sigma$ uncertainties on the flux (gray), along with the best fitting \texttt{alf} models used to derive the ages and elemental abundances (red) of four example primary Heavy Metal Survey targets (two at $z\sim1.4$ and two at $z\sim2.2$). The other spectra can be found in Appendix~\ref{app:spectra}. Spectra are binned such that one pixel is $\simeq5\,\mathrm{\mathring{A}}$ in the rest-frame. The spectra were fit before binning.}
    \label{fig:spec1}
\end{figure*}

\section{Sample and Methods}
\label{sec:methods}

\subsection{The Heavy Metal Survey}


The targets in this study are selected from the Keck Heavy Metal Survey; an ultra-deep rest-frame optical spectroscopic survey of massive quiescent galaxies at $z=1.3-2.2$. The 21 quiescent targets were identified using the UltraVISTA $K$-band selected catalog (v4.1) by \citet{muzzin_evolution_2013} and were selected to be quiescent using $UVJ$ colors \citep{wuyts_evolution_2016,williams_detection_2009}. The redshift intervals \lowzr~and \highzr~were chosen such that key rest-frame NIR metal absorption features, namely Balmer lines, H$\beta$, and MgI, fall within the wavelength windows of the Keck/LRIS and Keck/MOSFIRE spectrometers. Primary targets were further required to have $J<21.5$ and $H<21.7$ at \lowz\ and \highz, respectively

The $z\sim1.4$ targets were observed for a total of $\sim$17\;hrs on Keck ($\sim$5\;hrs with LRIS using the 600/10000 red grating and $\sim$12\;hrs with MOSFIRE-$J$), while the $z\sim2.1$ targets were observed for $\sim$30\;hrs ($\sim$13\;hrs with MOSFIRE-$J$ and $\sim$16\;hrs with MOSFIRE-$H$). These integration times provide sufficient S/N for modeling the faint stellar absorption features of individual massive quiescent galaxies. The typical S/N for the $z\sim1.4$ sample is 8$\,\mathrm{\mathring{A}^{-1}}$ and 15$\,\mathrm{\mathring{A}^{-1}}$ in the rest-frame for LRIS and MOSFIRE-J, respectively, while the S/N for the $z\sim2.1$ sample is 4$\,\mathrm{\mathring{A}^{-1}}$ and 12$\,\mathrm{\mathring{A}^{-1}}$ for MOSFIRE-J and MOSFIRE-H. For further details on the target selection, observing strategy, data reduction, galaxy sample characteristics, and stellar population properties (i.e., stellar masses) we refer to \citet{kriek_heavy_2023}.

One of the 21 primary targets (ID:55878) is removed from this study before fitting due to strong line emission, likely originating from AGN (see Ma et al. 2023, in prep). Another galaxy is also removed (ID:59449) because there are no clear absorption lines in the spectra.

\subsection{Stellar ages and elemental abundances}
\label{sec:alf}

We derive chemical compositions and stellar ages of the 19 Heavy Metal targets using the full-spectrum absorption line fitter (\texttt{alf}) code \citep{conroy_counting_2012, conroy_metal-rich_2018}. \texttt{alf} is designed to fit the individual elemental abundances and stellar ages from optical-NIR stellar continuum of $>1$\,Gyr stellar populations. \texttt{alf} combines metallicity-dependent MIST isochrones \citep{choi_mesa_2016}, empirical MILES and IRTF spectral libraries \citep{sanchez-blazquez_medium-resolution_2006, villaume_extended_2017}, and synthetic metallicity- and age-dependent elemental response spectra for 19 elements. 

\section{Results}
\begin{deluxetable}{ccccccc}
\tabletypesize{\footnotesize}\tablecolumns{8}
\tablewidth{0pt}
\tablecaption{Heavy Metal Quiescent Galaxy Parameters \label{table:alf-results}}
\tablehead{\colhead{ID\tablenotemark{a}} & \colhead{$z$} & \colhead{$\sigma$} & \colhead{Age} & \colhead{[Fe/H]} & \colhead{[Mg/Fe]}\\ 
\colhead{} & \colhead{} & \colhead{(km/s)} & \colhead{(Gyr)} & \colhead{} & \colhead{}}
\startdata
23621 & 1.358 & $199^{+38}_{-41}$ & $2.1^{+1.0}_{-0.7}$ & $-0.41^{+0.36}_{-0.41}$ & $0.31^{+0.34}_{-0.37}$\\ 
25407 & 1.359 & $201^{+27}_{-28}$ & $2.5^{+0.9}_{-0.6}$ & $-0.47^{+0.26}_{-0.24}$ & $0.64^{+0.24}_{-0.26}$\\ 
26888 & 1.359 & $166^{+33}_{-27}$ & $3.7^{+2.3}_{-1.1}$ & $-0.10^{+0.20}_{-0.26}$ & $0.27^{+0.22}_{-0.23}$\\ 
217249\tablenotemark{b,c} & 1.377 & $185^{+21}_{-24}$ & $0.8^{+0.2}_{-0.1}$ & $-0.38^{+0.21}_{-0.26}$ & $0.34^{+0.25}_{-0.25}$\\ 
214695\tablenotemark{b} & 1.396 & $210^{+24}_{-25}$ & $3.4^{+2.2}_{-0.9}$ & $-0.29^{+0.24}_{-0.27}$ & $0.34^{+0.18}_{-0.19}$\\ 
213947\tablenotemark{b} & 1.397 & $216^{+16}_{-13}$ & $0.9^{+0.1}_{-0.1}$ & $-0.17^{+0.17}_{-0.18}$ & $0.38^{+0.20}_{-0.21}$\\ 
213931\tablenotemark{b} & 1.399 & $371^{+13}_{-13}$ & $2.6^{+0.6}_{-0.4}$ & $-0.12^{+0.13}_{-0.12}$ & $0.27^{+0.07}_{-0.07}$\\ 
214340\tablenotemark{b} & 1.418 & $149^{+19}_{-23}$ & $1.4^{+0.4}_{-0.3}$ & $-0.34^{+0.20}_{-0.20}$ & $0.24^{+0.23}_{-0.23}$\\ 
23482 & 1.419 & $289^{+38}_{-35}$ & $2.6^{+1.3}_{-0.7}$ & $-0.09^{+0.24}_{-0.26}$ & $0.29^{+0.23}_{-0.22}$\\ 
25702 & 1.419 & $243^{+21}_{-20}$ & $1.8^{+0.7}_{-0.4}$ & $-0.24^{+0.18}_{-0.17}$ & $0.32^{+0.15}_{-0.16}$\\ 
23351 & 1.420 & $257^{+14}_{-13}$ & $1.3^{+0.3}_{-0.2}$ & $-0.07^{+0.15}_{-0.17}$ & $0.12^{+0.14}_{-0.13}$\\ 
59375\tablenotemark{c} & 1.550 & $333^{+42}_{-43}$ & $9.5^{+3.4}_{-4.9}$ & $0.21^{+0.30}_{-0.30}$ & $0.36^{+0.28}_{-0.29}$\\ 
60736\tablenotemark{c} & 1.861 & $270^{+30}_{-27}$ & $2.1^{+1.7}_{-0.6}$ & $-0.49^{+0.23}_{-0.25}$ & $0.60^{+0.26}_{-0.30}$\\ 
104779 & 2.123 & $186^{+30}_{-31}$ & $1.5^{+0.9}_{-0.4}$ & $-0.15^{+0.24}_{-0.28}$ & $0.55^{+0.23}_{-0.23}$\\ 
56163\tablenotemark{c} & 2.160 & $213^{+54}_{-48}$ & $2.9^{+1.1}_{-1.1}$ & $-0.43^{+0.32}_{-0.32}$ & $0.29^{+0.32}_{-0.31}$\\ 
106812 & 2.230 & $187^{+41}_{-45}$ & $1.5^{+1.1}_{-0.4}$ & $-0.44^{+0.28}_{-0.28}$ & $0.41^{+0.36}_{-0.32}$\\ 
108899 & 2.233 & $306^{+24}_{-23}$ & $1.2^{+0.2}_{-0.1}$ & $-0.38^{+0.22}_{-0.30}$ & $0.48^{+0.25}_{-0.29}$\\ 
107590\tablenotemark{c} & 2.234 & $173^{+25}_{-23}$ & $0.8^{+0.1}_{-0.1}$ & $-0.42^{+0.18}_{-0.22}$ & $0.44^{+0.26}_{-0.30}$\\ 
103236 & 2.243 & $197^{+37}_{-35}$ & $1.5^{+1.0}_{-0.4}$ & $-0.23^{+0.22}_{-0.23}$ & $0.50^{+0.25}_{-0.23}$\\ 
\enddata
\vspace{0.1cm}
\tablenotemark{a}{See \citep{kriek_heavy_2023} for coordinates, magnitudes, and stellar population and structural properties}
\vspace{-0.1cm}
\tablenotetext{b}{Appears in \citet{kriek_stellar_2019}}
\vspace{-0.1cm}
\tablenotetext{c}{Poorly constrained, removed from analysis}
\vspace{-1cm}
\end{deluxetable}

In the fitting presented here, we assume a model with a single burst of star formation (approximated as an SSP) and a fixed \citet{kroupa_variation_2001} IMF. We note that \texttt{alf} also allows for a young second SSP component to account for late-time star formation. In the end, the full-spectrum model has 26 free parameters: velocity offset, velocity dispersion, SSP age, isochrone metallicity, the abundances of 19 individual elements (Fe, O, C, N, Na, Mg, Si, K, Ca, Ti, V, Cr, Mn, Co, Ni, Cu, Sr, Ba, and Eu), and the emission line strengths of H, [OIII], and [NII].

The fitting is done in the rest-frame and the instrumental resolution of the data is accounted for by convolving the grid of models to match the resolution of the spectrum. We measure this instrumental resolution for each spectrum by finding the average full width at half maximum (FWHM) of 20-25 skylines in the variance spectrum. The average instrumental dispersions are $\sigma_{\rm inst}=$62, 36, and 33 km/s for the LRIS-Red, MOSFIRE-J, and MOSFIRE-H spectra, respectively. We note that the instrumental dispersion is lower than the template resolution in \texttt{alf}. This template resolution can be an issue if the intrinsic velocity dispersion is lower than 100 km/s. However, as we will find, the intrinsic velocity dispersions of the Heavy Metal galaxies are significantly higher than 100 km/s. The spectral continuum is also removed from the observations by fitting a high-order polynomial to the ratio of the data to the model (where $n \equiv (\lambda_{\rm max} - \lambda_{\rm min})/100\,{\rm \AA}$). We test the impact of the polynomial on the results by assuming different orders than the default ($n=8$ and $n=6$). We find that the polynomial order does not make a meaningful impact on the results, to well within the $1\sigma$ uncertainties. Furthermore, \citet{conroy_metal-rich_2018} test the impact of the normalizing polynomial on the fitting results and find that if fluxing issues are approximately linear on $<50\,\AA$ scales, the default polynomial assumption is sufficient. Once normalized, the spectrum is fit using Markov-Chain Monte Carlo \citep{foreman-mackey_emcee_2013} with 1,024 walkers and a burn-in of 20,000 steps. The priors of the MCMC fit are uniform and the walkers are initialized around the solar abundance pattern. 

After fitting all 19 galaxies, we visually inspect the best-fit models and the corresponding posterior distribution functions (PDFs) and remove any galaxy with poorly constrained parameters. We removed three galaxies because their PDFs do not follow normal distributions and/or they have ill-defined peaks. Two of these galaxies lack spectral coverage of key absorption features (D4000 break), leading to unreliable abundance measurements, and the third galaxy has strong [O\,{\sc ii}] and Hydrogen emission. We also remove two galaxies with best-fit stellar ages $<1$\,Gyr, as the stellar population models are not optimized for these young ages. As a sanity check, we also inspect correlations between all fitted parameters to ensure the relevant parameters are not being driven by other poorly constrained parameters. After removing these five galaxies, we are left with elemental abundances and stellar age measurements for ten galaxies at $z\sim1.4$ and four galaxies at $z\sim2.1$.


Five of the original 19 galaxies in this study were originally presented in \citet{kriek_stellar_2019}. In this study, we have since re-extracted the 1D spectra using an improved optimal weighing procedure. We also assume a single SSP when fitting, whereas \citet{kriek_stellar_2019} assume a two-burst model. Nonetheless, the results presented here are consistent to within 1$\sigma$ of what is found in \citet{kriek_stellar_2019}.

Example spectra with corresponding best-fit models are shown in Figure~\ref{fig:spec1}, while the remaining spectra and fits can be found in Appendix~\ref{app:spectra}. Corner plots corresponding to the galaxies in Figure~\ref{fig:spec1} are included in Appendix~\ref{app:corner_plots}. The best-fit parameters are listed in Table~\ref{table:alf-results} for all 19 galaxies. Objects with poorly constrained abundances are noted and are removed from the following analysis. In Appendix~\ref{app:recovery}, we conduct a mock recovery test to check for systematic uncertainties in the \texttt{alf} fitting as a function of SNR. Systematic uncertainties remain $<$0.1\;dex even at the lowest S/N (8\AA$^{-1}$). We refer to \citet{kriek_heavy_2023} for tabulated details of the observations and SED fitting results. 

In addition to elemental abundances and stellar ages, the full-spectrum fitting provides measurements of the stellar velocity dispersions. We find that the velocity dispersions of the Heavy Metal galaxies agree with previous measurements of the dynamical mass-stellar mass relation at similar redshifts \citep[e.g.,][]{belli_mosfire_2016} and we explore the implications of the Heavy Metal dynamical masses in the companion survey paper \citep{kriek_heavy_2023}.


\begin{figure*}
    \centering
    \includegraphics[width=\textwidth]{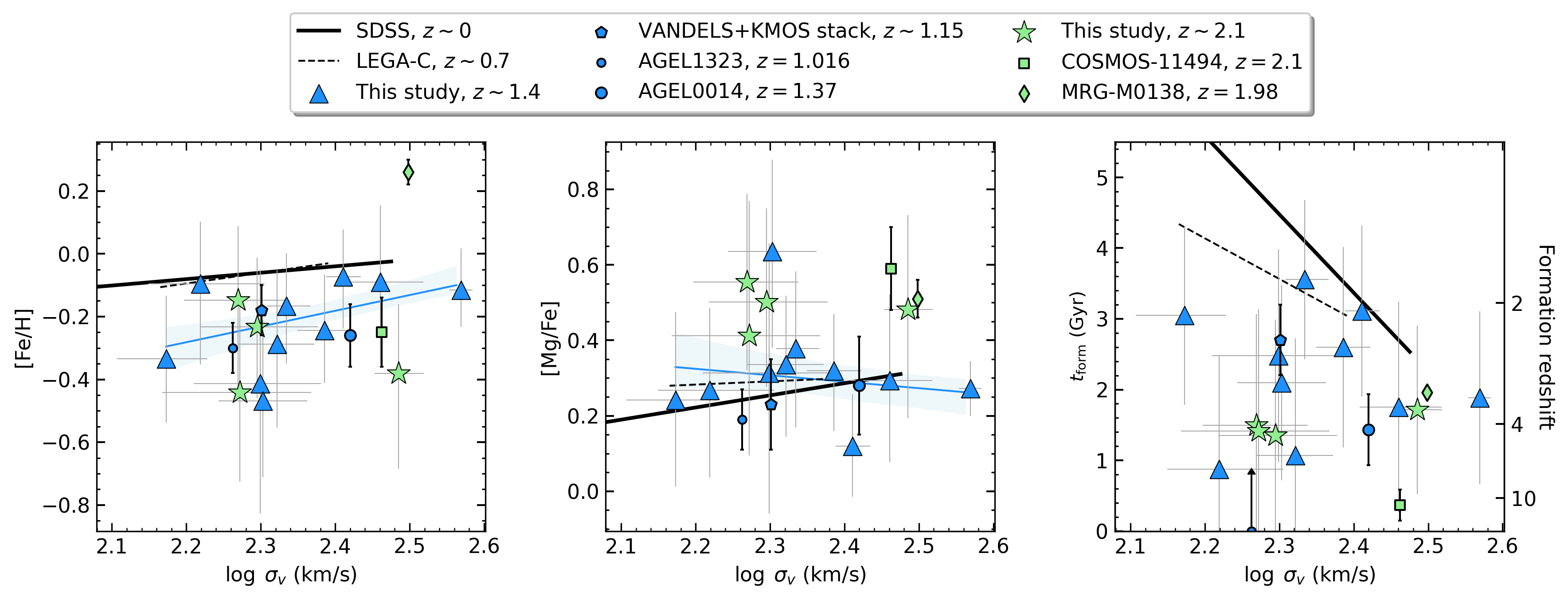}
    \caption{[Fe/H] (left), [Mg/Fe] (middle), and formation redshift (right) results for the $z\sim1.4$ (blue triangles) and $z\sim2.1$ (green stars) Heavy Metal targets as a function of stellar velocity dispersion. The black solid line shows the $\sigma_v$-abundance relations of massive quiescent galaxies at $z\sim0$ \citep{conroy_early-type_2014,beverage_elemental_2021}, while the black dashed lines denote relations at $z\sim0.7$ \citep{beverage_carbon_2023}. We include a quiescent galaxy at $z=2.1$ \citep[green square;][]{kriek_massive_2016}, a lensed quiescent galaxy at $z=1.98$ \citep[green diamond;][]{jafariyazani_resolved_2020}, a stack of quiescent galaxies at $1.0<z<1.3$ \citep[blue pentagon;][]{carnall_stellar_2022}, and two lensed quiescent galaxies at $z=1.016$ and $z=1.37$ \citep[blue circles;][]{zhuang_glimpse_2023}. The solid blue line in the left panel represents the best-fit $\sigma_v$-[Fe/H] with the 16th and 84th percentile uncertainties on the fit shown in light blue. The results from all studies were derived using \texttt{alf}. Quiescent galaxies at $z>0.7$ have lower [Fe/H], enhanced [Mg/Fe], and earlier formation times than galaxies at lower redshifts.}
    \label{fig:results1}
\end{figure*}

Figure~\ref{fig:results1} shows the variation in [Fe/H], [Mg/Fe], and formation time ($t_\mathrm{form}$) as a function of velocity dispersion $\sigma_v$ for the Heavy Metal galaxies at $z\sim1.4$ (blue triangles) and $z\sim2.1$ (purple stars). For comparison, we also show trends of quiescent galaxies at $z\sim0$ and $z\sim0.7$ from \citet{beverage_carbon_2023} in Figure~\ref{fig:results1}. The $z\sim0$ trends are derived by re-fitting the stacked spectra of a sample of 5,234 quiescent galaxies in SDSS from \citet{conroy_early-type_2014}, while the $z\sim0.7$ trends are based on 135 quiescent galaxies in the LEGA-C survey. Initially, \citet{beverage_elemental_2021} reported a -0.2\,dex offset in [Fe/H] between LEGA-C and SDSS galaxies based on the spectra from the second LEGA-C data release (DR2). However, with the subsequent release, DR3, which doubled the sample size and improved data reduction \citep[see][]{van_der_wel_large_2021}, \citet{beverage_carbon_2023} no longer found an offset in [Fe/H] between $z\sim0.7$ and $z\sim0$. Figure~\ref{fig:results1} also includes results from other studies of quiescent galaxies at $z\sim1.5$ \citep{carnall_stellar_2022, zhuang_glimpse_2023} and $z\sim2.1$ \citep{kriek_massive_2016, jafariyazani_resolved_2020}.
\footnote{\citet{carnall_stellar_2022} do not measure a velocity dispersion. We infer one using the average stellar mass of galaxies in the stacked spectrum and an empirically calibrated $\sigma_v$-$M_*$ relation from \citet{zahid_scaling_2016}.} 
We note that several other works have derived stellar metallicities of distant quiescent galaxies by fitting low-resolution grism spectra \citep[e.g.,][]{morishita_metal_2018,estrada-carpenter_clear_2019}. However, such studies rely on the shape of the stellar continuum to constrain the metallicities, leading to high degeneracies with stellar age, and therefore we do not include these results in our comparison. To ensure unbiased comparisons, we restrict the analysis to studies that utilize \texttt{alf}.

In the left panel of Figure~\ref{fig:results1} we show that the Heavy Metal galaxies have lower [Fe/H] compared to the $z\sim0$ and $z\sim0.7$ galaxies with similar velocity dispersions. Both the $z\sim1.4$ and $z\sim2.1$ galaxy populations have Fe-abundances between -0.5 and -0.1\;dex, with typical uncertainties of 0.2\;dex. There is a hint that the $z\sim2.1$ are slightly offset to lower [Fe/H] compared to $z\sim1.4$. We fit a $\sigma$-[Fe/H] relation to the $z\sim1.4$ population using a simple linear regression, with confidence intervals determined by perturbing the [Fe/H] of each data point according to their uncertainties and re-fitting 1000 times. The resulting relation is shown in the left panel of Figure~\ref{fig:results1} (blue line, with shaded confidence intervals). We find a positive slope of $0.52^{+0.38}_{-0.23}$, which is consistent with the $z\sim0$ and $z\sim0.7$ relations. However, there is a clear offset to lower [Fe/H]. We do not fit a relation to the $z\sim2.1$ population because of the small sample size. 

The observed evolution in [Fe/H] between $z=1.4-2.1$ and $z<1$ is corroborated by studies of other quiescent galaxies at similar redshifts. At $z\sim1.4$, galaxies in the VANDELS+KMOS survey \citep{carnall_stellar_2022} and two lensed galaxies in the AGEL survey \citep{zhuang_glimpse_2023} have [Fe/H]$\sim-0.25$, in agreement with the $z\sim1.4$ Heavy Metal galaxies (see Figure~\ref{fig:results1}). At $z=2.1$, \citet{kriek_massive_2016} measure [Fe/H]$=-0.25$ for a massive quiescent galaxy (COSMOS-11494), consistent with the values of the $z\sim2.1$ Heavy Metal galaxies (refer to Figure~\ref{fig:results1}). Interestingly, the only other study at $z\sim2.1$ finds super-solar [Fe/H] for a lensed galaxy (MRG-M0138) at $z=1.98$ \citep{jafariyazani_resolved_2020}. It was previously unclear whether this large variation in [Fe/H] is typical of the $z\sim2.1$ quiescent galaxy population. Now, with four Heavy Metal galaxies at $z\sim2.1$ with sub-solar [Fe/H], there is additional evidence suggesting that typical massive galaxies at $z\sim2.1$ are indeed Fe-deficient and that the Fe-abundance of MRG-M0138 may be anomalous. Nevertheless, the sample size is small and the large uncertainties on the Fe-abundances prevent any definitive statement about the $z\sim2.1$ population.

\begin{figure*}
  \centering
  \includegraphics[width=1\textwidth]{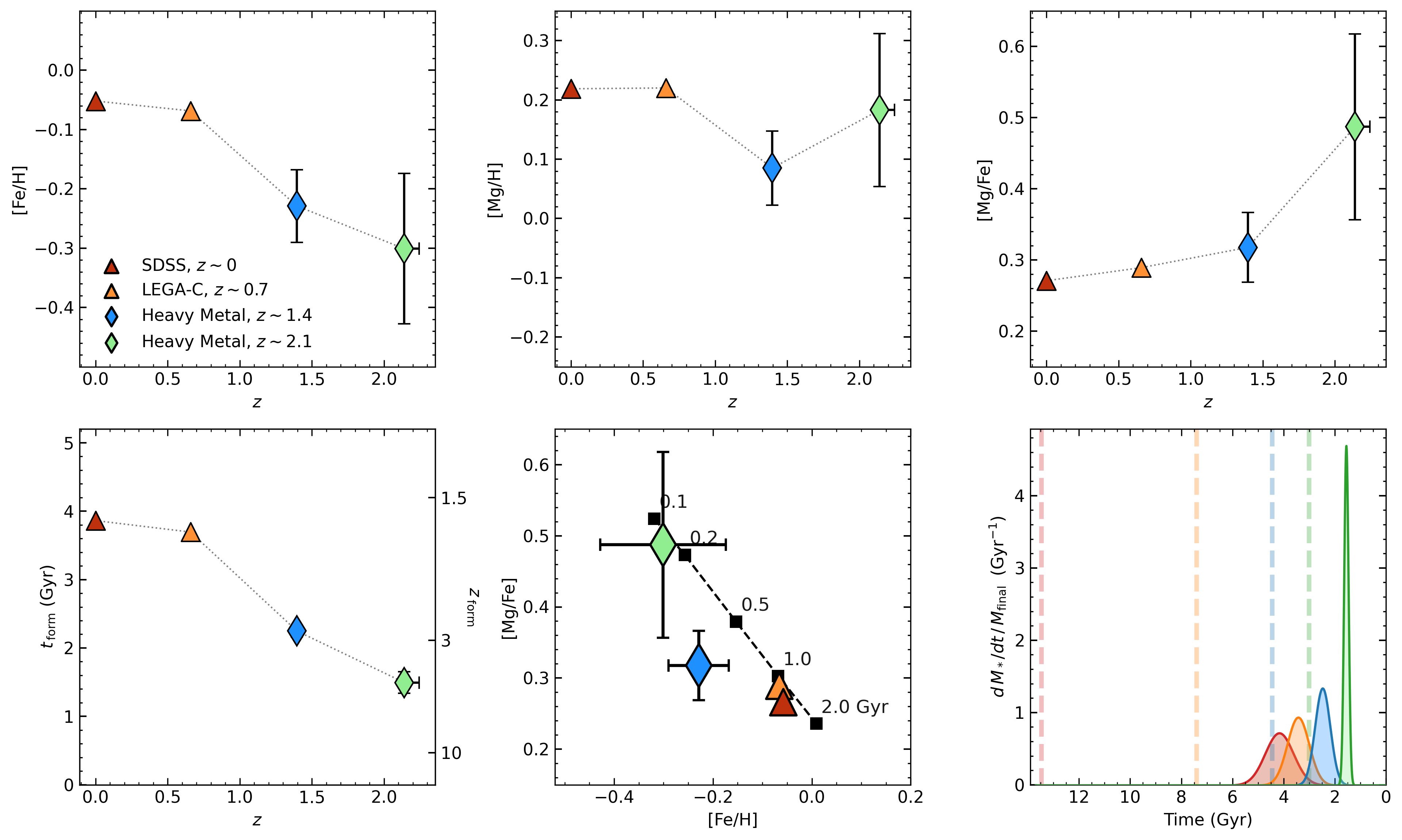} \\
  \vspace{0.1cm}
  \caption{Average elemental abundance and stellar age results for the Heavy Metal galaxies at $z\sim1.4$ and $z\sim2.1$, and for stacks of massive quiescent galaxies at $z\sim0$ and $z\sim0.7$ across various axes. The error bars on the SDSS and LEGA-C data points are smaller than the symbol size. \textit{Top row:} the [Fe/H], [Mg/H], and [Mg/Fe] abundances as a function of redshift. \textit{Bottom row:} the left panel shows the formation time as a function of redshift. In the middle panel, we compare the average abundances on the [Fe/H]-[Mg/Fe] plane to a simple closed-box chemical evolution model (black dashed line). Black squares represent predicted elemental abundances for various star-formation timescales, denoted in Gyr. The right panel shows the inferred star-formation history at each redshift interval. The age of the universe for each sample is marked with a dashed line. Typical quiescent galaxies at higher redshift formed earlier and over shorter timescales than massive quiescent galaxies today.}
  \label{fig:evolution}
  \vspace{0.2cm}
\end{figure*}

In the middle panel of Figure~\ref{fig:results1} we see that the Heavy Metal galaxies at $z\sim1.4$ have similar [Mg/Fe] to what is found in the SDSS ($z\sim0$) and LEGA-C ($z\sim0.7$) samples. Meanwhile, the $z\sim2.1$ Heavy Metal galaxies have [Mg/Fe] that are $\approx$0.2\;dex higher than what is found at lower redshifts, though typical uncertainties are large (0.2\,dex) and the sample size is small. We derive a $\sigma$-[Mg/Fe] relation for the $z\sim1.4$ Heavy Metal galaxies using the same method as with [Fe/H], resulting in a negative slope of $-0.15^{+0.17}_{-0.29}$ (blue line with shaded confidence intervals). This slope is consistent with zero to within 1$\sigma$ and there is no significant offset compared to the $z<1$ relations. These results are supported by the other studies of quiescent galaxies at $z>1$. The two studies at $z\sim1.4$ find [Mg/Fe] consistent with the LEGA-C and SDSS, while the two at $z\sim2$ find enhanced [Mg/Fe] consistent with the $z\sim2.1$ Heavy Metal galaxies.

Finally, in the right panel of Figure~\ref{fig:results1} we show galaxy $\sigma_v$ vs. galaxy formation time. This age is determined using the best-fit stellar population age. We find that the formation time of the Heavy Metal galaxies range from 1\;Gyr ($z_\mathrm{form}=9$) and 4\;Gyr ($z_\mathrm{form}=2$) after the Big Bang, with an average formation time of 2.5\;Gyr ($z_\mathrm{form}=3$) and 1.5\;Gyr ($z_\mathrm{form}=4$) for the $z\sim1.4$ and $z\sim2.1$ samples, respectively. We note that the galaxy formation times are calculated using an SSP-equivalent age, and more complex SFHs would have likely resulted in slightly earlier formation times. The assumption of an SSP, however, does not bias the stellar metallicities or elemental abundance measurements \citep{serra_interpretation_2007}. In the next section we consider the results in Figure~\ref{fig:results1} in the context of galaxy evolution.


\section{The evolution of massive quiescent galaxy population since \texorpdfstring{$\lowercase{z}=2.1$}{z=2.1}}

In this section, we investigate the changes of \textit{average} stellar metallicities, abundance ratios, and ages of massive quiescent galaxies as a function of redshift and discuss implications for their evolution over the past 11 billion years. In Figure~\ref{fig:evolution} we show the average [Fe/H], [Mg/H], $t_\mathrm{\,form}$, and [Mg/Fe] for quiescent galaxies in the SDSS, LEGA-C, and Heavy Metal surveys as a function of redshift. \textbf{To allow for a fair comparison between surveys, we only consider the SDSS and LEGA-C galaxies that fall in bins with similar $\sigma_v$ as the Heavy Metal galaxies. Specifically, we remove three SDSS bins and one LEGA-C bin with $\sigma<170$~km/s.  Thus, the average velocity dispersions of the SDSS and LEGA-C galaxies used in Figure~\ref{fig:evolution} are $\sigma_v=240\,{\rm km\, s^{-1}}$, with the number of galaxies in each respective sample now being 2,990 (SDSS) and 101 (LEGA-C).}

Examining the top left panel of Figure~\ref{fig:evolution}, we observe an increase in the average [Fe/H] from $z=2.1$ to $z=0$, with distinct offsets of -0.2 and -0.3\;dex at $z\sim1.4$ and $z\sim2.1$, respectively, between the Heavy Metal galaxies and the lower redshift samples. Moving to the middle upper panel, we find less evolution in [Mg/H] compared to [Fe/H]; the $z=1.4$ sample is 0.1 dex lower compared to the $z=0.7$ and $z=0$ galaxies. The $z=2.1$ galaxies have no [Mg/H] offset compared to lower redshift galaxies, although uncertainties remain substantial. Primarily due to the increase in [Fe/H], the average [Mg/Fe] decreases towards lower redshift, as revealed in the upper right panel. The $z=2.1$ sample exhibits [Mg/Fe] elevated by 0.25 dex, while the $z=1.4$ sample is only elevated by 0.08 dex.
In the bottom left panel we show that the average $t_\mathrm{\,form}$ changes from $t_\mathrm{\,form}=1.5$ to $t_\mathrm{\,form}=4\,$Gyr from $z=2.1$ to $z=0$. 

In the bottom middle panel of Figure~\ref{fig:evolution} we compare the average [Mg/Fe] as a function of [Fe/H] with a simple closed-box chemical evolution model from \citet[][represented by a black dashed line]{kriek_massive_2016}. The model provides mass-weighted abundances for various different star-formation timescales (indicated in Gyr next to the black boxes). The behavior of the model is driven by the differential enrichment timescales of Fe and Mg. While Mg is predominantly synthesized in massive stars and promptly returned to the interstellar medium (ISM) through core-collapse supernovae (SNe), Fe is produced through the explosions of intermediate-mass stars (Type Ia SNe; SNIa) and is released on delayed timescales. Therefore, galaxies with shorter star-formation timescales experience less Fe-enrichment from SNIa prior to quenching, resulting in stellar populations with lower [Fe/H] and higher [Mg/Fe].

The simple chemical evolution model used in this paper is illustrative, and is by no means meant to capture the complex feedback processes and star-formation histories likely present in the galaxies at hand. More detailed chemical evolution models, including inflows and outflows, along with varied star-formation histories, will help decipher the various factors shaping the elemental abundances and stellar metallicities of integrated stellar populations, especially as we move to higher redshifts (e.g., see N. Gountanis in prep.).

A visual comparison between the observations and the simple chemical evolution model highlights that the inferred star-formation timescales of the $z\sim0$ and $z\sim0.7$ samples are longer than the extreme timescales observed for the Heavy Metal galaxies, which are approximately 0.15\;Gyr and 0.7\;Gyr for the $z\sim2.1$ and $z\sim1.4$ samples, respectively. These inferred star-formation timescales derived from [Mg/Fe], along with the results for $t_\mathrm{form}$, paint a picture of galaxy formation in which typical massive quiescent galaxies at higher redshift formed earlier and underwent more rapid star formation compared to their counterparts at $z\sim0$. To visually depict this scenario, we present the implied average star-formation histories in the lower right panel of Figure~\ref{fig:evolution}. We adopt a Gaussian parameterization for the star-formation histories, with the mean centered at $t_\mathrm{\,form}$ and the FWHM corresponding to the star-formation timescale, which is determined by comparing the abundances to the chemical evolution model. It is important to note that this panel serves as an illustrative tool, and the choice of a Gaussian star-formation history is driven by its simplicity rather than a specific physical motivation. Nevertheless, it is evident that the star-formation histories of the Heavy Metal galaxies, particularly at $z\sim2.1$, are extremely short compared to the lower redshift samples. We also note that this panel does not take into account the evolution of the stellar mass function, nor the spread in formation redshifts and star-formation timescales within the sample. Instead, it shows the \textit{typical} SFH at each redshift interval.

Before considering how one should interpret the observed evolution, it is important to understand how selection biases influence the results. Unlike the Heavy Metal survey, the $z\sim0$ and $z\sim0.7$ samples are mass-complete, and therefore, the average stellar population properties should reflect the general galaxy population at each redshift. The Heavy Metal sample, however, has a complex selection function \citep[see][]{kriek_heavy_2023}. The selection favors quiescent galaxies with lower mass-to-light ratios (M/L), indicative of younger ages. As a result, the ages of the Heavy Metal galaxies are likely biased young compared to the true population of quiescent galaxies at z$>$1. Accounting for this bias, however, would strengthen the observed evolution in Figure~\ref{fig:evolution}; galaxies at z$>$1 could have even earlier formation redshifts, and therefore more extreme chemical properties, than what is found in this sample.

One possible explanation for the observed evolution is the growth of the quiescent galaxy population due to the continuous quenching of star-forming galaxies between $z=2.1$ and $z=0$ \citep[i.e., progenitor bias;][]{van_dokkum_morphological_2001,carollo_newly_2013,lilly_gas_2013}. Galaxies that quench at later epochs presumably form their stars over longer timescales, resulting in later $t_\mathrm{\,form}$ and more Fe-enrichment from SNIa (lower [Mg/Fe]) before quenching. Hence, the addition of these galaxies to the quiescent population steadily increases the \textit{average} $t_\mathrm{form}$ and [Fe/H], as observed. In fact, the number of quiescent galaxies with $M_*>10^{10.5}\;M_\odot$ has increased tenfold since $z=2$, as evident from the evolution of the stellar mass function \citep[e.g.,][]{conselice_direct_2022, mcleod_evolution_2021}. Furthermore, most of this evolution occurs at $z>0.5$ \citep{mcleod_evolution_2021}, which could explain the pronounced jump in [Fe/H] and $t_\mathrm{\,form}$ from $z\sim1.4$ to $z\sim0.7$.

An alternate explanation is mergers and late-time star formation, which impact the elemental abundances of \textit{individual} galaxies; when galaxies accrete satellites or form new stars, their integrated elemental abundances are altered by the newly added stars. Indeed, the importance of (dry) minor mergers in the evolution of quiescent galaxies has been posited by observations that they were more compact and had flatter color gradients at earlier times \citep[e.g.,][]{van_dokkum_confirmation_2008, van_dokkum_growth_2010, naab_minor_2009,trujillo_superdense_2009,taylor_masses_2010,suess_half-mass_2019-1,suess_color_2020, miller_color_2023} as well as by direct pair fraction measurements \citep{newman_can_2012}. Furthermore, \citet{suess_minor_2023} show that tiny companions with stellar mass ratios 1:900 compromise some 30\% of their total stellar mass budget.

The extent to which mergers, late-time star formation, and progenitor bias contribute to the evolution of the massive quiescent galaxy population remains a topic of ongoing debate \citep[e.g.,][]{bezanson_relation_2009, hopkins_compact_2009, carollo_newly_2013, poggianti_evolution_2013, van_de_sande_stellar_2013, williams_morphology_2017, damjanov_quiescent_2019}. In the following section we examine the elemental abundances and stellar ages of individual galaxies to help differentiate between the various evolutionary scenarios; if mergers and/or late-time star formation play only a small role in the evolution of massive quiescent galaxies, then the extreme galaxies observed at $z\sim2$ should persist as rare remnants in the $z\sim0$ population.

\begin{figure*}
    \centering
    \includegraphics[width=\textwidth]{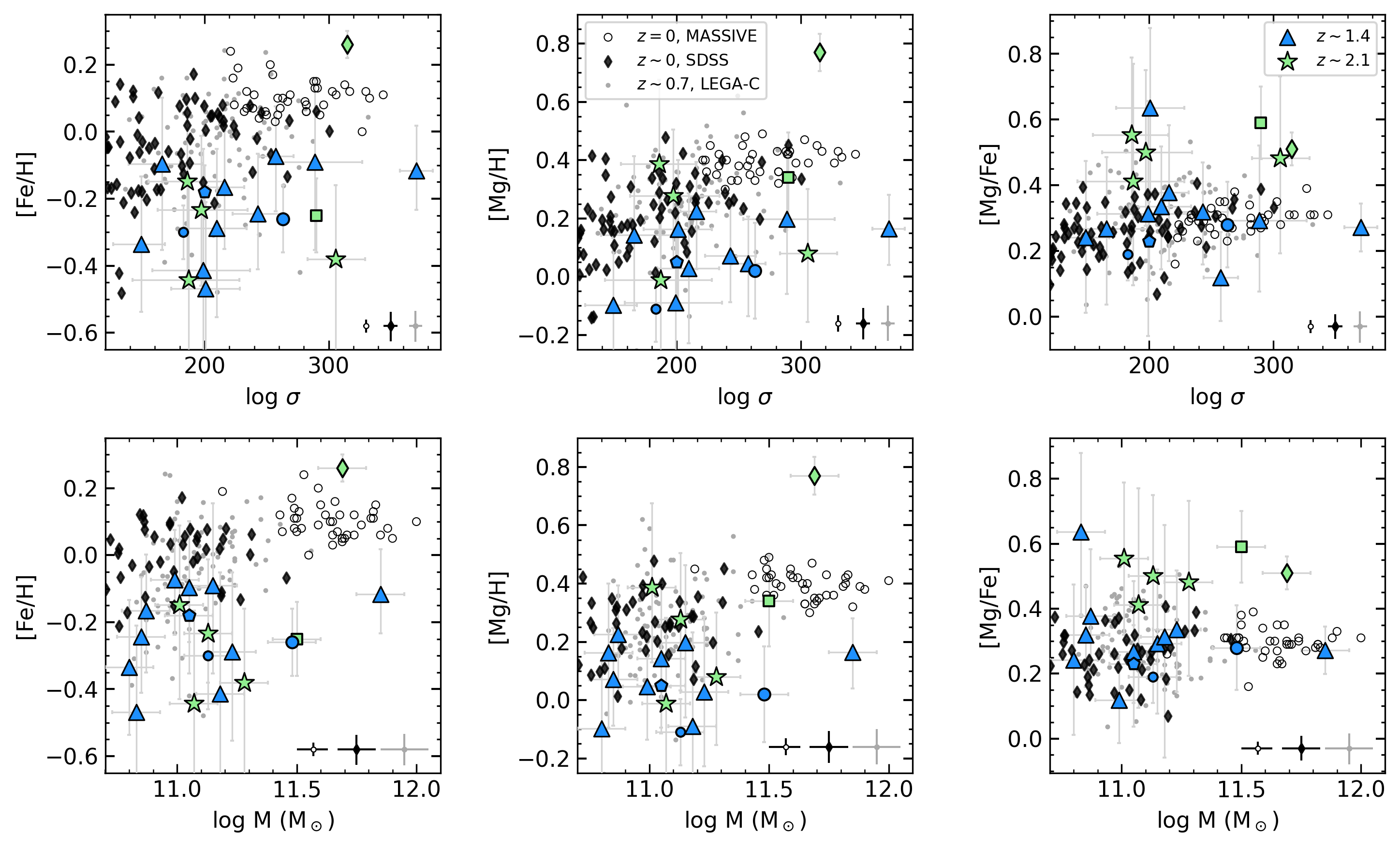}
    \caption{Individual measurements of [Fe/H], [Mg/H], and [Mg/Fe] as a function of $\sigma_v$ (top row) and $M_*$ (bottom row). The Heavy Metal galaxies are split into $z\sim1.4$ (blue triangles) and $z\sim2.1$ (green stars) samples. We include results from other studies of massive quiescent galaxies at similar redshifts (see legend in Figure~\ref{fig:results1}). We compare these results to individual abundance measurements from the MASSIVE \citep[$z=0$;][open circles]{gu_massive_2022}, SDSS \citep[$z\sim0$;][black diamonds]{zhuang_glimpse_2023}, and LEGA-C \citep[$z\sim0.7$;][gray dots]{beverage_carbon_2023} surveys, with their typical uncertainties provided at the bottom of each panel. At constant $M_*$ and $\sigma_v$, galaxies at $z>1$ have different chemical properties than those at lower redshifts.}
    \label{fig:massive}
\end{figure*}

\section{Implications for the build-up of massive quiescent galaxies}

\begin{figure}
    \centering
    \includegraphics[width=\columnwidth]{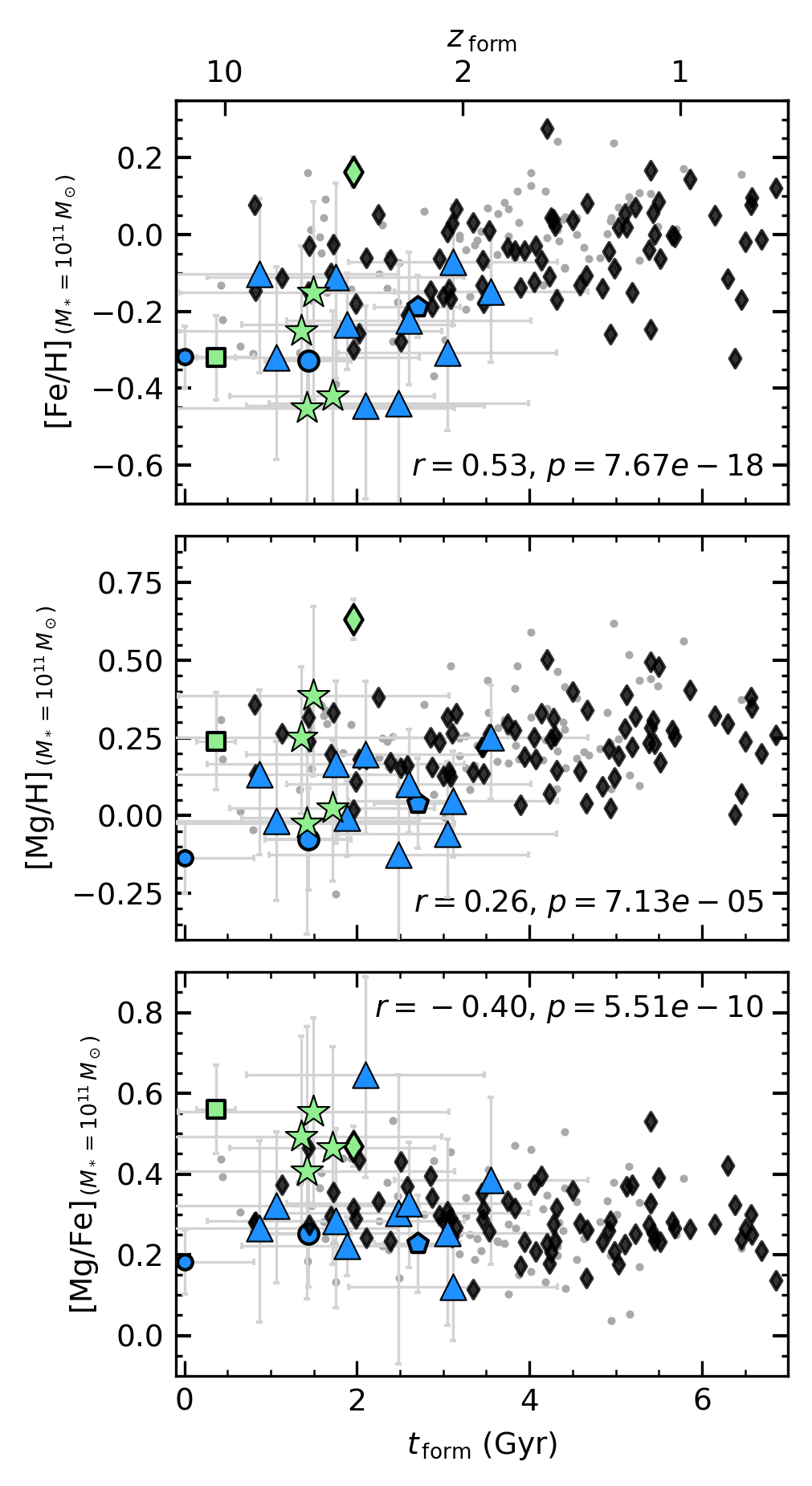}
    \caption{[Fe/H], [Mg/H], and [Mg/Fe] at a fixed stellar mass (${M_*=10^{11}M_\odot}$) as a function of formation time for the LEGA-C (gray circles), Heavy Metal (blue triangles and green stars), SDSS (red shading) and various other $z>1$ studies (refer to the legend in Figure~\ref{fig:results1}). Abundances at \textit{fixed} stellar mass are calculated from the $M_*$-abundance relations at $z\sim0$ shown in Figure~\ref{fig:massive}. We assess the strength of the correlation between the elemental abundances and formation time using the Pearson correlation test (results listed in each panel). Galaxies that form at later cosmic epochs have higher [Fe/H] and [Mg/H], and slightly lower [Mg/Fe]. At a fixed $t_{\rm\,form}$, the Heavy Metal galaxies exhibit similar abundances to the SDSS and LEGA-C populations.}
    \label{fig:form_v_abund}
\end{figure}

In the previous section, we observed significant evolution in the average elemental abundances of massive quiescent galaxies from $z\sim2.1$ to $z\sim0$, suggesting that galaxies at higher redshift are typically metal-poor, have shorter star-formation timescales, and formed earlier. In this section, we examine the abundance ratios of individual galaxies over cosmic time to help differentiate between various evolutionary channels.

Figure~\ref{fig:massive} shows [Fe/H], [Mg/H], and [Mg/Fe] as a function of $\sigma_v$ (top row) and $M_*$ (bottom row). The stellar masses of the Heavy Metal galaxies are derived in \citep{kriek_heavy_2023}. We compare the Heavy Metal galaxies to the abundances of individual LEGA-C galaxies \citep[gray circles,]{beverage_carbon_2023}, and to galaxies in SDSS (black diamonds) presented by \citet{zhuang_glimpse_2023}. Additionally, we compare to \texttt{alf}-derived abundances of 41 galaxies from the MASSIVE survey, a volume-limited survey of the most massive galaxies in the local universe \citep[black diamonds,][]{ma_massive_2014,gu_massive_2022}. We also include abundance results from other studies of quiescent galaxies at $z\sim1.4$ and $z\sim2.1$, with the same markers as Figure~\ref{fig:results1}. We note that the SDSS fiber, and the LEGA-C and Heavy Metal slits cover similar physical extents within the galaxies (3-4 kpc). However, as galaxies were smaller at earlier times, the Heavy Metal and LEGA-C slits cover 1-1.2 $R_e$, while SDSS only covers 0.4-0.8 $R_e$. The MASSIVE spectra were extracted from the inner $R_e/8$ and therefore represent only the cores of nearby massive early-type galaxies. 



In the previous section, we proposed two theories to explain the observed abundance differences between the Heavy Metal survey and $z\sim0$. In the first scenario, the quiescent galaxy population is continuously growing by the addition of newly quenched galaxies, driving the evolution of the observed average abundances \citep[``progenitor bias," e.g.,][]{van_dokkum_morphological_2001, carollo_newly_2013, poggianti_evolution_2013}. In this scenario, we expect to find galaxies with properties similar to the Heavy Metal galaxies in the lower redshift universe, and thus the populations at higher and lower redshift should overlap. In other words, the Heavy Metal galaxies should constitute the lower metallicity and high-[Mg/Fe] tail of the $z\sim0$ samples. Indeed, the top panels of Figure~\ref{fig:massive} show that this is mostly the case for the Heavy Metal galaxies, though there are a few exceptions; most of the Heavy Metal galaxies with high $\sigma_v$ have significantly lower [Fe/H] and [Mg/H] than any galaxy found at $z\sim0$ and $z\sim0.7$. Furthermore, in the rightmost panel, we find that galaxies at $z\sim2.1$ are \textit{all} [Mg/Fe]-enhanced compared to the galaxies at lower redshifts. 

When considered as a function of stellar mass, the metallicity differences between the lower and higher redshift populations become more apparent. At all stellar masses, and particularly at the highest stellar masses, the Heavy Metal galaxies have [Fe/H] and [Mg/H] $\approx0.2\;$dex lower than any individual galaxy in SDSS or LEGA-C. This result poses a problem for the progenitor bias scenario; if these Heavy Metal galaxies passively evolve to $z\sim0$, we would expect to find Fe- and Mg-deficient remnants in the SDSS and MASSIVE populations. Specifically, in this scenario, 10\% of the SDSS and MASSIVE galaxies should be direct descendants of $z>2$ quiescent galaxies given the rapid evolution of the stellar mass function \citep{mcleod_evolution_2021}. Thus, the $z\sim0$ galaxies are likely not the direct descendants of the $z\sim2$ Heavy Metal galaxies. It is important to acknowledge, however, that uncertainties are substantial and sample sizes are limited. Thus, the possibility remains that we may have simply missed the rare metal-poor remnants in the $z\sim0$ samples. We also note that the Heavy Metal galaxies were selected to maximize the number of bright quiescent galaxies in one field of view. As such, the environments of these galaxies may not be representative of the broader massive quiescent galaxy population at higher redshift.

As previously mentioned, minor mergers provide an alternate way to explain the observed evolution. In this picture, the $z\sim1.4-2.1$ quiescent galaxies may grow into the cores of today's massive early-type galaxies by accreting low-mass galaxies \citep[e.g.,][]{bezanson_relation_2009, hopkins_compact_2009, van_de_sande_stellar_2013}. This scenario can successfully explain the observed increase in stellar metallicity and decrease in [Mg/Fe], but only if the accreted galaxies form over long timescales (i.e. lower [Mg/Fe]) and are more metal-rich. There are, however, two problems with this scenario. Firstly, our observations show there is evolution within the inner $1~R_e$, while minor mergers are typically accreted onto galaxy outskirts \citep{naab_minor_2009, oser_two_2010}. Secondly, given the mass-metallicity relation, the accreted low-mass galaxies will likely be metal-poor, and therefore cannot explain the increase in metallicity \citep[][]{kirby_universal_2013}. Major mergers with galaxies that have longer star-formation timescales (and thus lower [Mg/Fe]) solve both of these problems, as more massive galaxies tend to be more metal-rich \textit{and} can more easily disrupt galaxy cores. However, the observed major merger rate is likely too low to fully account for the observed evolution \citep{oser_cosmological_2012, nipoti_size_2012, newman_can_2012} and major mergers alone cannot explain the size evolution \citep[e.g.,][]{hopkins_compact_2009, van_dokkum_growth_2010, hilz_how_2013, belli_velocity_2014}. Therefore, a combination of major and minor mergers may be necessary to explain the observed evolution. 

Finally, late-time star formation, likely in combination with or triggered by mergers, can increase the metallicities of individual galaxies while decreasing [Mg/Fe]; if the star-formation material is enriched in SNIa products by previous epochs of star formation, then the newly formed stars will be younger, more metal-rich, and more $\alpha$-enhanced. It should be noted that a burst of central star-formation is expected in a wet merger scenario \citep[e.g.,][]{hopkins_unified_2006} and thus the two scenarios together are effective at explaining the absence of metal-poor and [Mg/Fe]-enriched galaxies by $z\sim0$ as observed in Figure~\ref{fig:massive}. 

Crucially, all of the above scenarios rely on the assumption that at the same stellar masses, younger galaxies have higher metallicity and lower [Mg/Fe]. For example, the progenitor bias scenario only works if the galaxies that quench at later times are more metal-rich and form over longer timescales. In Figure~\ref{fig:form_v_abund} we search for such trends in [Fe/H], [Mg/H], and [Mg/Fe] as a function of $t_{\rm\,form}$ for the SDSS, LEGA-C, and Heavy Metal quiescent galaxies. In each panel, we remove the first-order mass dependence by subtracting the SDSS $M_*$-abundance relations from all galaxies and then scaling them to the value of the $M_*$-abundance relations at $M_*=10^{11}\,M_\odot$ \citep[e.g.,][]{leethochawalit_evolution_2019,zhuang_glimpse_2023}. To assess the level of correlation between quenching time (as traced by $t_{\rm form}$) and stellar metallicities, we use the Pearson correlation coefficient. A correlation is deemed ``marginal'' if the Pearson correlation coefficient is $0.2\leq r< 0.4$, ``moderate'' if $0.4\leq r<0.6$, and ``strong'' if $r\geq 0.6$. We list the correlation coefficients along with the associated p-value in Figure~\ref{fig:form_v_abund}. At fixed stellar mass, we do in fact find that younger SDSS and LEGA-C galaxies exhibit higher [Fe/H] ($r=0.53$), marginally higher [Mg/H] ($r=0.26$), and moderately lower [Mg/Fe] ($r=-0.40$), as previously found by \citet{beverage_carbon_2023} and \citet{zhuang_glimpse_2023}, respectively \citep[see also][]{leethochawalit_evolution_2018}.

The trends with [Fe/H] and [Mg/Fe] can be naturally explained by galaxies joining the quiescent galaxy population at later times (higher $t_{\rm\,form}$) having more time to form stars from gas that is rich in SNIa products (namely, Fe). However, the marginal trend with [Mg/H] is more challenging to explain, as Mg-enrichment is independent of the star-formation timescale and instead primarily depends on the size of the gas reservoir at the time of star-formation quenching. \citet{beverage_elemental_2021} propose a model to explain the marginal trend with $t_{\rm \, form}$ by invoking rapid gas removal during the quenching process. In summary, if galaxies at earlier epochs, that form earlier and over shorter timescales,  quench more abruptly and have larger gas fractions at the time of quenching (e.g. effective AGN outflows), then [Mg/H] is truncated to lower values. Thus, the marginal trend with [Mg/H] may suggest that galaxies at earlier times experience more effective outflows \citep[see also the outflow model proposed by][]{leethochawalit_evolution_2019}. 

Furthermore, Figure~\ref{fig:form_v_abund} shows that high-$z$ galaxies have a slight systematic offset to lower metallicities compared to local galaxies even when the dependence on $M_*$ and stellar age is removed. This same result was also found by \citet{zhuang_glimpse_2023}, but for a smaller galaxy sample. Thus, this result confirms the finding by \citet{zhuang_glimpse_2023} and reinforces the need for mergers or late-time star formation to alter the chemical compositions and stellar ages of these extreme individual galaxies by $z\sim0$. Again, we note that comparisons at ``fixed formation times'' between galaxies at different redshift regimes should be approached with caution; the SSP-equivalent ages can be biased young. As a result, a more complicated star-formation history may shift the Heavy Metal galaxies in Figure~\ref{fig:form_v_abund} towards even earlier formation times. As a result, the ages of the Heavy Metal galaxies are likely biased young compared to the true population of quiescent galaxies at z$>$1. Nonetheless, correcting for such a bias would not change our interpretation.

Figures~\ref{fig:massive} and \ref{fig:form_v_abund} together demonstrate that progenitor bias, minor and major mergers, and late-time star formation are likely all responsible, to some extent, for the observed evolution in the elemental abundances of massive quiescent galaxies over cosmic time. This analysis further shows that the existing story of massive quiescent galaxy assembly -- namely, that the $z>1$ quiescent galaxies are the progenitor cores of today's massive early-type galaxies -- is more complicated than initially anticipated. We arrive at a similar conclusion based on the dynamical masses of the Heavy Metal galaxies in \citet{kriek_heavy_2023}.

\section{Summary and Conclusion}

In this paper, we present the stellar metallicities, elemental abundance ratios, and stellar ages of 19 massive quiescent galaxies from the Keck Heavy Metal survey. The galaxies are divided over two redshift intervals around $z\sim1.4$ and $z\sim2.1$. We fit the ultra-deep LRIS+MOSFIRE spectra using a full-spectrum modeling code with variable elemental abundance ratios. Combined, these measurements represent the largest sample of individual quiescent galaxies at $z\gtrsim1.4$ with stellar metallicity and elemental abundance measurements.

We find that massive quiescent galaxies at $1.4\lesssim z\lesssim 2.2$ have subsolar Fe-abundances, ranging from [Fe/H]$\;=-0.5$ to $-0.1$~dex, with typical values of $-0.2$ and $-0.3$~dex at $z\sim1.4$ and $z\sim2.1$, respectively. We also find a trend between $\sigma_v$ and [Fe/H], with a best-fit slope consistent with what is found at lower redshifts. We find high [Mg/Fe] of $\approx0.5$ for the $z\sim2.1$ galaxies, while those at $z\sim1.4$ are lower with [Mg/Fe]$\;\sim0.3$. The ages range from 0.9--3.7~Gyr, implying formation redshifts $z_{\rm\,form}=2-8$. 

We examine the evolution of the massive quiescent galaxy population over cosmic time by comparing the average Heavy Metal elemental abundances and ages to those in the LEGA-C ($z\sim0.7$) and SDSS ($z\sim0$) surveys. The $z\sim1.4$ and $z\sim2.1$ galaxies have [Fe/H] that are $0.15$ and $0.23$~dex lower than what is observed at $z\lesssim0.7$. Interestingly, we find no evolution in [Mg/Fe] at $z\lesssim1.4$, though the $z\sim2.1$ galaxies have 0.2~dex higher [Mg/Fe]. By comparing these results with a simple chemical evolution model, we find that galaxies at higher redshifts formed at progressively earlier epochs and shorter timescales, with the $z\sim2.1$ galaxies forming in only $\approx150$\;Myr at $z=4$. 

The observed evolution of stellar population properties provides a unique insight into the assembly histories of massive quiescent galaxies. One possible explanation is the growth of the quiescent galaxy population. If the galaxies that join the population at later times are younger, more metal-rich, and less alpha-enhanced, then the average properties will evolve. However, this scenario cannot fully explain the observations because we do not find the direct descendants of the chemically extreme Heavy Metal galaxies at $z\lesssim0.7$. Minor mergers also fail to account for the observed evolution; due to the mass-metallicity relation, they cannot increase [Fe/H], in particular in the galaxy centers. Thus, major mergers and/or late-time star formation are required to explain the evolution of chemical properties across cosmic time. These results imply a more complicated assembly history than the minor-merger driven inside-out growth scenario previously inferred from the evolution in the sizes, color gradients, and central densities of quiescent galaxies. In our accompanying paper, \citet{kriek_heavy_2023}, arrive at a similar conclusion based on the dynamical masses of the Heavy Metal galaxies.

This study was enabled by ultra-deep spectroscopic observations conducted with the most efficient ground-based spectrographs. With up to 16\;hr of integration per galaxy per band, these spectra represent some of the deepest spectroscopic observations of individual galaxies at $z\gtrsim1.4$ with coverage of key absorption features. As a result, the Heavy Metal survey has tripled the number of individual stellar metallicity measurements at $z\gtrsim1.4$. Nonetheless, sample sizes at these redshifts remain small, and the uncertainties on each measurement remain substantial. 

In the future, JWST will provide even deeper spectroscopy of massive quiescent galaxies at $z\gtrsim1.4$, extending the existing sample sizes, reaching to lower stellar masses, and enabling the measurements of more elements. Thus, JWST will revolutionize our understanding of the assembly of massive quiescent galaxies over comsic time. \\

We would like to thank Zhuyun Zhuang and Glenn Van de Ven for useful conversations. We acknowledge support from NSF AAG grants AST-1908748 and 1909942. A.G.B. is supported by the National Science Foundation Graduate Research Fellowship Program under grant No. DGE 1752814 and DGE 2146752 and by the H2H8 Association and LKBF. We gratefully acknowledge the efforts and dedication of the Keck Observatory staff for support, especially amid the challenges posed by a global pandemic. We recognize and acknowledge the very significant cultural role and reverence that the summit of Mauna Kea has within the indigenous Hawaiian community. We are deeply fortunate to have the opportunity to conduct observations from this mountain. 

\software{\texttt{alf} \citep{conroy_counting_2012,conroy_metal-rich_2018}, \texttt{FAST} \citep{kriek_ultra-deep_2009}}

\clearpage

\bibliography{zoterolib}

\begin{thebibliography}{}
\expandafter\ifx\csname natexlab\endcsname\relax\def\natexlab#1{#1}\fi
\providecommand{\url}[1]{\href{#1}{#1}}
\providecommand{\dodoi}[1]{doi:~\href{http://doi.org/#1}{\nolinkurl{#1}}}
\providecommand{\doeprint}[1]{\href{http://ascl.net/#1}{\nolinkurl{http://ascl.net/#1}}}
\providecommand{\doarXiv}[1]{\href{https://arxiv.org/abs/#1}{\nolinkurl{https://arxiv.org/abs/#1}}}

\bibitem[{Antwi-Danso {et~al.}(2023)Antwi-Danso, Papovich, Esdaile,
  Nanayakkara, Glazebrook, Hutchison, Whitaker, Marsan, Diaz, Marchesini,
  Muzzin, Tran, Setton, Kaushal, Speagle, \& Cole}]{antwi-danso_feniks_2023}
Antwi-Danso, J., Papovich, C., Esdaile, J., {et~al.} 2023, The {FENIKS}
  {Survey}: {Spectroscopic} {Confirmation} of {Massive} {Quiescent} {Galaxies}
  at z {\textasciitilde} 3-5,  arXiv.
\newblock \url{http://arxiv.org/abs/2307.09590}

\bibitem[{Barro {et~al.}(2017)Barro, Faber, Koo, Dekel, Fang, Trump,
  Pérez-González, Pacifici, Primack, Somerville, Yan, Guo, Liu, Ceverino,
  Kocevski, \& McGrath}]{barro_structural_2017}
Barro, G., Faber, S.~M., Koo, D.~C., {et~al.} 2017, The Astrophysical Journal,
  840, 47, \dodoi{10.3847/1538-4357/aa6b05}

\bibitem[{Belli {et~al.}(2014)Belli, Newman, \& Ellis}]{belli_velocity_2014}
Belli, S., Newman, A.~B., \& Ellis, R.~S. 2014, The Astrophysical Journal, 783,
  117, \dodoi{10.1088/0004-637X/783/2/117}

\bibitem[{Belli {et~al.}(2016)Belli, Newman, \& Ellis}]{belli_mosfire_2016}
---. 2016, The Astrophysical Journal, 834, 18,
  \dodoi{10.3847/1538-4357/834/1/18}

\bibitem[{Bevacqua {et~al.}(2023)Bevacqua, Saracco, La Barbera, D’Ago,
  De Propris, Ferreras, Gallazzi, Pasquali, \&
  Spiniello}]{bevacqua_elemental_2023}
Bevacqua, D., Saracco, P., La Barbera, F., {et~al.} 2023, Monthly Notices of
  the Royal Astronomical Society, 525, 4219, \dodoi{10.1093/mnras/stad2403}

\bibitem[{Beverage {et~al.}(2021)Beverage, Kriek, Conroy, Bezanson, Franx, \&
  van~der Wel}]{beverage_elemental_2021}
Beverage, A.~G., Kriek, M., Conroy, C., {et~al.} 2021, The Astrophysical
  Journal Letters, 917, L1, \dodoi{10.3847/2041-8213/ac12cd}

\bibitem[{Beverage {et~al.}(2023)Beverage, Kriek, Conroy, Sandford, Bezanson,
  Franx, Wel, \& Weisz}]{beverage_carbon_2023}
---. 2023, The Astrophysical Journal, 948, 140,
  \dodoi{10.3847/1538-4357/acc176}

\bibitem[{Bezanson {et~al.}(2009)Bezanson, van Dokkum, Tal, Marchesini, Kriek,
  Franx, \& Coppi}]{bezanson_relation_2009}
Bezanson, R., van Dokkum, P.~G., Tal, T., {et~al.} 2009, The Astrophysical
  Journal, 697, 1290, \dodoi{10.1088/0004-637X/697/2/1290}

\bibitem[{Carnall(2017)}]{carnall_spectres_2017}
Carnall, A.~C. 2017, {SpectRes}: {A} {Fast} {Spectral} {Resampling} {Tool} in
  {Python},  arXiv.
\newblock \url{http://arxiv.org/abs/1705.05165}

\bibitem[{Carnall {et~al.}(2022)Carnall, McLure, Dunlop, Hamadouche, Cullen,
  McLeod, Begley, Amorin, Bolzonella, Castellano, Cimatti, Fontanot, Gargiulo,
  Garilli, Mannucci, Pentericci, Talia, Zamorani, Calabro, Cresci, \&
  Hathi}]{carnall_stellar_2022}
Carnall, A.~C., McLure, R.~J., Dunlop, J.~S., {et~al.} 2022, The Astrophysical
  Journal, 929, 131, \dodoi{10.3847/1538-4357/ac5b62}

\bibitem[{Carnall {et~al.}(2023)Carnall, McLure, Dunlop, McLeod, Wild, Cullen,
  Magee, Begley, Cimatti, Donnan, Hamadouche, Jewell, \&
  Walker}]{carnall_massive_2023}
---. 2023, Nature, 619, 716, \dodoi{10.1038/s41586-023-06158-6}

\bibitem[{Carollo {et~al.}(2013)Carollo, Bschorr, Renzini, Lilly, Capak,
  Cibinel, Ilbert, Onodera, Scoville, Cameron, Mobasher, Sanders, \&
  Taniguchi}]{carollo_newly_2013}
Carollo, C.~M., Bschorr, T.~J., Renzini, A., {et~al.} 2013, The Astrophysical
  Journal, 773, 112, \dodoi{10.1088/0004-637X/773/2/112}

\bibitem[{Choi {et~al.}(2014)Choi, Conroy, Moustakas, Graves, Holden, Brodwin,
  Brown, \& van Dokkum}]{choi_assembly_2014}
Choi, J., Conroy, C., Moustakas, J., {et~al.} 2014, The Astrophysical Journal,
  792, 95, \dodoi{10.1088/0004-637X/792/2/95}

\bibitem[{Choi {et~al.}(2016)Choi, Dotter, Conroy, Cantiello, Paxton, \&
  Johnson}]{choi_mesa_2016}
Choi, J., Dotter, A., Conroy, C., {et~al.} 2016, The Astrophysical Journal,
  823, 102, \dodoi{10.3847/0004-637X/823/2/102}

\bibitem[{Cimatti {et~al.}(2004)Cimatti, Daddi, Renzini, Cassata, Vanzella,
  Pozzetti, Cristiani, Fontana, Rodighiero, Mignoli, \&
  Zamorani}]{cimatti_old_2004}
Cimatti, A., Daddi, E., Renzini, A., {et~al.} 2004, Nature, 430, 184,
  \dodoi{10.1038/nature02668}

\bibitem[{Cimatti {et~al.}(2008)Cimatti, Cassata, Pozzetti, Kurk, Mignoli,
  Renzini, Daddi, Bolzonella, Brusa, Rodighiero, Dickinson, Franceschini,
  Zamorani, Berta, Rosati, \& Halliday}]{cimatti_gmass_2008}
Cimatti, A., Cassata, P., Pozzetti, L., {et~al.} 2008, Astronomy \&
  Astrophysics, 482, 21, \dodoi{10.1051/0004-6361:20078739}

\bibitem[{Conroy {et~al.}(2014)Conroy, Graves, \& van
  Dokkum}]{conroy_early-type_2014}
Conroy, C., Graves, G.~J., \& van Dokkum, P.~G. 2014, The Astrophysical
  Journal, 780, 33, \dodoi{10.1088/0004-637X/780/1/33}

\bibitem[{Conroy \& van Dokkum(2012)}]{conroy_counting_2012}
Conroy, C., \& van Dokkum, P. 2012, The Astrophysical Journal, 747, 69,
  \dodoi{10.1088/0004-637X/747/1/69}

\bibitem[{Conroy {et~al.}(2018)Conroy, Villaume, van Dokkum, \&
  Lind}]{conroy_metal-rich_2018}
Conroy, C., Villaume, A., van Dokkum, P., \& Lind, K. 2018, The Astrophysical
  Journal, 854, 139, \dodoi{10.3847/1538-4357/aaab49}

\bibitem[{Conselice {et~al.}(2022)Conselice, Mundy, Ferreira, \&
  Duncan}]{conselice_direct_2022}
Conselice, C.~J., Mundy, C.~J., Ferreira, L., \& Duncan, K. 2022, The
  Astrophysical Journal, 940, 168, \dodoi{10.3847/1538-4357/ac9b1a}

\bibitem[{Daddi {et~al.}(2005)Daddi, Renzini, Pirzkal, Cimatti, Malhotra,
  Stiavelli, Xu, Pasquali, Rhoads, Brusa, Alighieri, Ferguson, Koekemoer,
  Moustakas, Panagia, \& Windhorst}]{daddi_passively_2005}
Daddi, E., Renzini, A., Pirzkal, N., {et~al.} 2005, The Astrophysical Journal,
  626, 680, \dodoi{10.1086/430104}

\bibitem[{Damjanov {et~al.}(2019)Damjanov, Zahid, Geller, Utsumi, Sohn, \&
  Souchereau}]{damjanov_quiescent_2019}
Damjanov, I., Zahid, H.~J., Geller, M.~J., {et~al.} 2019, The Astrophysical
  Journal, 872, 91, \dodoi{10.3847/1538-4357/aaf97d}

\bibitem[{Damjanov {et~al.}(2009)Damjanov, McCarthy, Abraham, Glazebrook, Yan,
  Mentuch, Borgne, Savaglio, Crampton, Murowinski, Juneau, Carlberg,
  Jørgensen, Roth, Chen, \& Marzke}]{damjanov_red_2009}
Damjanov, I., McCarthy, P.~J., Abraham, R.~G., {et~al.} 2009, The Astrophysical
  Journal, 695, 101, \dodoi{10.1088/0004-637X/695/1/101}

\bibitem[{Davies {et~al.}(1993)Davies, Sadler, \&
  Peletier}]{davies_line-strength_1993}
Davies, R.~L., Sadler, E.~M., \& Peletier, R.~F. 1993, Monthly Notices of the
  Royal Astronomical Society, 262, 650, \dodoi{10.1093/mnras/262.3.650}

\bibitem[{Dokkum {et~al.}(2008)Dokkum, Franx, Kriek, Holden, Illingworth,
  Magee, Bouwens, Marchesini, Quadri, Rudnick, Taylor, \&
  Toft}]{dokkum_confirmation_2008}
Dokkum, P. G.~v., Franx, M., Kriek, M., {et~al.} 2008, The Astrophysical
  Journal, 677, L5, \dodoi{10.1086/587874}

\bibitem[{Estrada-Carpenter {et~al.}(2019)Estrada-Carpenter, Papovich,
  Momcheva, Brammer, Long, Quadri, Bridge, Dickinson, Ferguson, Finkelstein,
  Giavalisco, Gosmeyer, Lotz, Salmon, Skelton, Trump, \&
  Weiner}]{estrada-carpenter_clear_2019}
Estrada-Carpenter, V., Papovich, C., Momcheva, I., {et~al.} 2019, The
  Astrophysical Journal, 870, 133, \dodoi{10.3847/1538-4357/aaf22e}

\bibitem[{Foreman-Mackey {et~al.}(2013)Foreman-Mackey, Hogg, Lang, \&
  Goodman}]{foreman-mackey_emcee_2013}
Foreman-Mackey, D., Hogg, D.~W., Lang, D., \& Goodman, J. 2013, Publications of
  the Astronomical Society of the Pacific, 125, 306, \dodoi{10.1086/670067}

\bibitem[{Franx {et~al.}(2003)Franx, Labbé, Rudnick, Dokkum, Daddi, Schreiber,
  Moorwood, Rix, Röttgering, Wel, Werf, \&
  Starkenburg}]{franx_significant_2003}
Franx, M., Labbé, I., Rudnick, G., {et~al.} 2003, The Astrophysical Journal,
  587, L79, \dodoi{10.1086/375155}

\bibitem[{Gallazzi {et~al.}(2005)Gallazzi, Charlot, Brinchmann, White, \&
  Tremonti}]{gallazzi_ages_2005}
Gallazzi, A., Charlot, S., Brinchmann, J., White, S. D.~M., \& Tremonti, C.~A.
  2005, Monthly Notices of the Royal Astronomical Society, 362, 41,
  \dodoi{10.1111/j.1365-2966.2005.09321.x}

\bibitem[{Glazebrook {et~al.}(2017)Glazebrook, Schreiber, Labbé, Nanayakkara,
  Kacprzak, Oesch, Papovich, Spitler, Straatman, Tran, \&
  Yuan}]{glazebrook_massive_2017}
Glazebrook, K., Schreiber, C., Labbé, I., {et~al.} 2017, Nature, 544, 71,
  \dodoi{10.1038/nature21680}

\bibitem[{Gu {et~al.}(2022)Gu, Greene, Newman, Kreisch, Quenneville, Ma, \&
  Blakeslee}]{gu_massive_2022}
Gu, M., Greene, J.~E., Newman, A.~B., {et~al.} 2022, The Astrophysical Journal,
  932, 103, \dodoi{10.3847/1538-4357/ac69ea}

\bibitem[{Hilz {et~al.}(2013)Hilz, Naab, \& Ostriker}]{hilz_how_2013}
Hilz, M., Naab, T., \& Ostriker, J.~P. 2013, Monthly Notices of the Royal
  Astronomical Society, 429, 2924, \dodoi{10.1093/mnras/sts501}

\bibitem[{Hopkins {et~al.}(2009)Hopkins, Bundy, Murray, Quataert, Lauer, \&
  Ma}]{hopkins_compact_2009}
Hopkins, P.~F., Bundy, K., Murray, N., {et~al.} 2009, Monthly Notices of the
  Royal Astronomical Society, 398, 898,
  \dodoi{10.1111/j.1365-2966.2009.15062.x}

\bibitem[{Hopkins {et~al.}(2006)Hopkins, Hernquist, Cox, Matteo, Robertson, \&
  Springel}]{hopkins_unified_2006}
Hopkins, P.~F., Hernquist, L., Cox, T.~J., {et~al.} 2006, The Astrophysical
  Journal Supplement Series, 163, 1, \dodoi{10.1086/499298}

\bibitem[{Jafariyazani {et~al.}(2020)Jafariyazani, Newman, Mobasher, Belli,
  Ellis, \& Patel}]{jafariyazani_resolved_2020}
Jafariyazani, M., Newman, A.~B., Mobasher, B., {et~al.} 2020, The Astrophysical
  Journal, 897, L42, \dodoi{10.3847/2041-8213/aba11c}

\bibitem[{Ji \& Giavalisco(2022)}]{ji_reconstructing_2022}
Ji, Z., \& Giavalisco, M. 2022, The Astrophysical Journal, 935, 120,
  \dodoi{10.3847/1538-4357/ac7f43}

\bibitem[{Khochfar \& Silk(2009)}]{khochfar_dry_2009}
Khochfar, S., \& Silk, J. 2009, Monthly Notices of the Royal Astronomical
  Society, 397, 506, \dodoi{10.1111/j.1365-2966.2009.14958.x}

\bibitem[{Kirby {et~al.}(2013)Kirby, Cohen, Guhathakurta, Cheng, Bullock, \&
  Gallazzi}]{kirby_universal_2013}
Kirby, E.~N., Cohen, J.~G., Guhathakurta, P., {et~al.} 2013, The Astrophysical
  Journal, 779, 102, \dodoi{10.1088/0004-637X/779/2/102}

\bibitem[{Kriek {et~al.}(2009)Kriek, van Dokkum, Labbé, Franx, Illingworth,
  Marchesini, \& Quadri}]{kriek_ultra-deep_2009}
Kriek, M., van Dokkum, P.~G., Labbé, I., {et~al.} 2009, The Astrophysical
  Journal, 700, 221, \dodoi{10.1088/0004-637X/700/1/221}

\bibitem[{Kriek {et~al.}(2006)Kriek, Dokkum, Franx, Quadri, Gawiser, Herrera,
  Illingworth, Labbé, Lira, Marchesini, Rix, Rudnick, Taylor, Toft, Urry, \&
  Wuyts}]{kriek_spectroscopic_2006}
Kriek, M., Dokkum, P. G.~v., Franx, M., {et~al.} 2006, The Astrophysical
  Journal, 649, L71, \dodoi{10.1086/508371}

\bibitem[{Kriek {et~al.}(2016)Kriek, Conroy, van Dokkum, Shapley, Choi, Reddy,
  Siana, van~de Voort, Coil, \& Mobasher}]{kriek_massive_2016}
Kriek, M., Conroy, C., van Dokkum, P.~G., {et~al.} 2016, Nature, 540, 248,
  \dodoi{10.1038/nature20570}

\bibitem[{Kriek {et~al.}(2019)Kriek, Price, Conroy, Suess, Mowla, Pasha,
  Bezanson, van Dokkum, \& Barro}]{kriek_stellar_2019}
Kriek, M., Price, S.~H., Conroy, C., {et~al.} 2019, The Astrophysical Journal,
  880, L31, \dodoi{10.3847/2041-8213/ab2e75}

\bibitem[{Kriek {et~al.}(2023)Kriek, Beverage, Price, Suess, Barro, Bezanson,
  Conroy, Cutler, Franx, Lin, Lorenz, Ma, Momcheva, Mowla, Pasha, van Dokkum,
  \& Whitaker}]{kriek_heavy_2023}
Kriek, M., Beverage, A.~G., Price, S.~H., {et~al.} 2023, The {Heavy} {Metal}
  {Survey}: {Star} {Formation} {Constraints} and {Dynamical} {Masses} of 21
  {Massive} {Quiescent} {Galaxies} at z{\textasciitilde}1.4-2.2,  arXiv.
\newblock \url{http://arxiv.org/abs/2311.16232}

\bibitem[{Kroupa(2001)}]{kroupa_variation_2001}
Kroupa, P. 2001, Monthly Notices of the Royal Astronomical Society, 322, 231,
  \dodoi{10.1046/j.1365-8711.2001.04022.x}

\bibitem[{Kuntschner(2004)}]{kuntschner_line--sight_2004}
Kuntschner, H. 2004, Astronomy \& Astrophysics, 426, 737,
  \dodoi{10.1051/0004-6361:20041414}

\bibitem[{Leethochawalit {et~al.}(2019)Leethochawalit, Kirby, Ellis, Moran, \&
  Treu}]{leethochawalit_evolution_2019}
Leethochawalit, N., Kirby, E.~N., Ellis, R.~S., Moran, S.~M., \& Treu, T. 2019,
  The Astrophysical Journal, 885, 100, \dodoi{10.3847/1538-4357/ab4809}

\bibitem[{Leethochawalit {et~al.}(2018)Leethochawalit, Kirby, Moran, Ellis, \&
  Treu}]{leethochawalit_evolution_2018}
Leethochawalit, N., Kirby, E.~N., Moran, S.~M., Ellis, R.~S., \& Treu, T. 2018,
  The Astrophysical Journal, 856, 15, \dodoi{10.3847/1538-4357/aab26a}

\bibitem[{Lilly {et~al.}(2013)Lilly, Carollo, Pipino, Renzini, \&
  Peng}]{lilly_gas_2013}
Lilly, S.~J., Carollo, C.~M., Pipino, A., Renzini, A., \& Peng, Y. 2013, The
  Astrophysical Journal, 772, 119, \dodoi{10.1088/0004-637X/772/2/119}

\bibitem[{Lonoce {et~al.}(2015)Lonoce, Longhetti, Maraston, Thomas, Mancini,
  Cimatti, Ciocca, Citro, Daddi, Alighieri, Gargiulo, Maiolino, Mannucci,
  Moresco, Pozzetti, Quai, \& Saracco}]{lonoce_old_2015}
Lonoce, I., Longhetti, M., Maraston, C., {et~al.} 2015, Monthly Notices of the
  Royal Astronomical Society, 454, 3912, \dodoi{10.1093/mnras/stv2150}

\bibitem[{Ma {et~al.}(2014)Ma, Greene, McConnell, Janish, Blakeslee, Thomas, \&
  Murphy}]{ma_massive_2014}
Ma, C.-P., Greene, J.~E., McConnell, N., {et~al.} 2014, The Astrophysical
  Journal, 795, 158, \dodoi{10.1088/0004-637X/795/2/158}

\bibitem[{Maiolino \& Mannucci(2019)}]{maiolino_re_2019}
Maiolino, R., \& Mannucci, F. 2019, The Astronomy and Astrophysics Review, 27,
  3, \dodoi{10.1007/s00159-018-0112-2}

\bibitem[{Matteucci(1994)}]{matteucci_abundance_1994}
Matteucci, F. 1994, Astronomy and Astrophysics, 288, 57.
\newblock \url{https://ui.adsabs.harvard.edu/abs/1994A&A...288...57M}

\bibitem[{McDermid {et~al.}(2015)McDermid, Alatalo, Blitz, Bournaud, Bureau,
  Cappellari, Crocker, Davies, Davis, de~Zeeuw, Duc, Emsellem, Khochfar,
  Krajnović, Kuntschner, Morganti, Naab, Oosterloo, Sarzi, Scott, Serra,
  Weijmans, \& Young}]{mcdermid_atlas3d_2015}
McDermid, R.~M., Alatalo, K., Blitz, L., {et~al.} 2015, Monthly Notices of the
  Royal Astronomical Society, 448, 3484, \dodoi{10.1093/mnras/stv105}

\bibitem[{McLean {et~al.}(2010)McLean, Steidel, Epps, Matthews, Adkins,
  Konidaris, Weber, Aliado, Brims, Canfield, Cromer, Fucik, Kulas, Mace,
  Magnone, Rodriguez, Wang, \& Weiss}]{mclean_design_2010}
McLean, I.~S., Steidel, C.~C., Epps, H., {et~al.} 2010, in Design and
  development of {MOSFIRE}: the multi-object spectrometer for infrared
  exploration at the {Keck} {Observatory}, ed. I.~S. McLean, S.~K. Ramsay, \&
  H.~Takami, San Diego, California, USA, 77351E, \dodoi{10.1117/12.856715}

\bibitem[{McLean {et~al.}(2012)McLean, Steidel, Epps, Konidaris, Matthews,
  Adkins, Aliado, Brims, Canfield, Cromer, Fucik, Kulas, Mace, Magnone,
  Rodriguez, Rudie, Trainor, Wang, Weber, \& Weiss}]{mclean_mosfire_2012}
McLean, I.~S., Steidel, C.~C., Epps, H.~W., {et~al.} 2012, in {MOSFIRE}, the
  multi-object spectrometer for infra-red exploration at the {Keck}
  {Observatory}, ed. I.~S. McLean, S.~K. Ramsay, \& H.~Takami, Amsterdam,
  Netherlands, 84460J, \dodoi{10.1117/12.924794}

\bibitem[{McLeod {et~al.}(2021)McLeod, McLure, Dunlop, Cullen, Carnall, \&
  Duncan}]{mcleod_evolution_2021}
McLeod, D.~J., McLure, R.~J., Dunlop, J.~S., {et~al.} 2021, Monthly Notices of
  the Royal Astronomical Society, 503, 4413, \dodoi{10.1093/mnras/stab731}

\bibitem[{Miller {et~al.}(2023)Miller, Dokkum, \& Mowla}]{miller_color_2023}
Miller, T.~B., Dokkum, P.~v., \& Mowla, L. 2023, The Astrophysical Journal,
  945, 155, \dodoi{10.3847/1538-4357/acbc74}

\bibitem[{Morishita {et~al.}(2018)Morishita, Abramson, Treu, Wang, Brammer,
  Kelly, Stiavelli, Jones, Schmidt, Trenti, \& Vulcani}]{morishita_metal_2018}
Morishita, T., Abramson, L.~E., Treu, T., {et~al.} 2018, The Astrophysical
  Journal, 856, L4, \dodoi{10.3847/2041-8213/aab493}

\bibitem[{Mowla {et~al.}(2019)Mowla, Dokkum, Brammer, Momcheva, Wel, Whitaker,
  Nelson, Bezanson, Muzzin, Franx, MacKenty, Leja, Kriek, \&
  Marchesini}]{mowla_cosmos-dash_2019}
Mowla, L.~A., Dokkum, P.~v., Brammer, G.~B., {et~al.} 2019, The Astrophysical
  Journal, 880, 57, \dodoi{10.3847/1538-4357/ab290a}

\bibitem[{Muzzin {et~al.}(2013)Muzzin, Marchesini, Stefanon, Franx, McCracken,
  Milvang-Jensen, Dunlop, Fynbo, Brammer, Labbé, \& van
  Dokkum}]{muzzin_evolution_2013}
Muzzin, A., Marchesini, D., Stefanon, M., {et~al.} 2013, The Astrophysical
  Journal, 777, 18, \dodoi{10.1088/0004-637X/777/1/18}

\bibitem[{Naab {et~al.}(2009)Naab, Johansson, \& Ostriker}]{naab_minor_2009}
Naab, T., Johansson, P.~H., \& Ostriker, J.~P. 2009, The Astrophysical Journal,
  699, L178, \dodoi{10.1088/0004-637X/699/2/L178}

\bibitem[{Nanayakkara {et~al.}(2023)Nanayakkara, Glazebrook, Jacobs,
  Kawinwanichakij, Schreiber, Brammer, Esdaile, Kacprzak, Labbe, Lagos,
  Marchesini, Marsan, Oesch, Papovich, Remus, \&
  Tran}]{nanayakkara_population_2023}
Nanayakkara, T., Glazebrook, K., Jacobs, C., {et~al.} 2023, A population of
  faint, old, and massive quiescent galaxies at 3 {\textless} z {\textless} 4
  revealed by {JWST} {NIRSpec} {Spectroscopy},  arXiv.
\newblock \url{http://arxiv.org/abs/2212.11638}

\bibitem[{Newman {et~al.}(2012)Newman, Ellis, Bundy, \& Treu}]{newman_can_2012}
Newman, A.~B., Ellis, R.~S., Bundy, K., \& Treu, T. 2012, The Astrophysical
  Journal, 746, 162, \dodoi{10.1088/0004-637X/746/2/162}

\bibitem[{Nipoti {et~al.}(2012)Nipoti, Treu, Leauthaud, Bundy, Newman, \&
  Auger}]{nipoti_size_2012}
Nipoti, C., Treu, T., Leauthaud, A., {et~al.} 2012, Monthly Notices of the
  Royal Astronomical Society, 422, 1714,
  \dodoi{10.1111/j.1365-2966.2012.20749.x}

\bibitem[{Oke {et~al.}(1995)Oke, Cohen, Carr, Cromer, Dingizian, Harris,
  Labrecque, Lucinio, Schaal, Epps, \& Miller}]{oke_keck_1995}
Oke, J.~B., Cohen, J.~G., Carr, M., {et~al.} 1995, Publications of the
  Astronomical Society of the Pacific, 107, 375, \dodoi{10.1086/133562}

\bibitem[{Onodera {et~al.}(2015)Onodera, Carollo, Renzini, Cappellari, Mancini,
  Arimoto, Daddi, Gobat, Strazzullo, Tacchella, \& Yamada}]{onodera_ages_2015}
Onodera, M., Carollo, C.~M., Renzini, A., {et~al.} 2015, The Astrophysical
  Journal, 808, 161, \dodoi{10.1088/0004-637X/808/2/161}

\bibitem[{Oser {et~al.}(2012)Oser, Naab, Ostriker, \&
  Johansson}]{oser_cosmological_2012}
Oser, L., Naab, T., Ostriker, J.~P., \& Johansson, P.~H. 2012, The
  Astrophysical Journal, 744, 63, \dodoi{10.1088/0004-637X/744/1/63}

\bibitem[{Oser {et~al.}(2010)Oser, Ostriker, Naab, Johansson, \&
  Burkert}]{oser_two_2010}
Oser, L., Ostriker, J.~P., Naab, T., Johansson, P.~H., \& Burkert, A. 2010, The
  Astrophysical Journal, 725, 2312, \dodoi{10.1088/0004-637X/725/2/2312}

\bibitem[{Peng {et~al.}(2015)Peng, Maiolino, \&
  Cochrane}]{peng_strangulation_2015}
Peng, Y., Maiolino, R., \& Cochrane, R. 2015, Nature, 521, 192,
  \dodoi{10.1038/nature14439}

\bibitem[{Poggianti {et~al.}(2013)Poggianti, Moretti, Calvi, D'Onofrio,
  Valentinuzzi, Fritz, \& Renzini}]{poggianti_evolution_2013}
Poggianti, B.~M., Moretti, A., Calvi, R., {et~al.} 2013, The Astrophysical
  Journal, 777, 125, \dodoi{10.1088/0004-637X/777/2/125}

\bibitem[{Rockosi {et~al.}(2010)Rockosi, Stover, Kibrick, Lockwood, Peck,
  Cowley, Bolte, Adkins, Alcott, Allen, Brown, Cabak, Deich, Hilyard, Kassis,
  Lanclos, Lewis, Pfister, Phillips, Robinson, Saylor, Thompson, Ward, Wei, \&
  Wright}]{rockosi_low-resolution_2010}
Rockosi, C., Stover, R., Kibrick, R., {et~al.} 2010, in Society of
  {Photo}-{Optical} {Instrumentation} {Engineers} ({SPIE}) {Conference}
  {Series}, Vol. 7735, Ground-based and {Airborne} {Instrumentation} for
  {Astronomy} {III}, ed. I.~S. McLean, S.~K. Ramsay, \& H.~Takami, 77350R,
  \dodoi{10.1117/12.856818}

\bibitem[{Sanchez-Blazquez {et~al.}(2006)Sanchez-Blazquez, Peletier,
  Jimenez-Vicente, Cardiel, Cenarro, Falcon-Barroso, Gorgas, Selam, \&
  Vazdekis}]{sanchez-blazquez_medium-resolution_2006}
Sanchez-Blazquez, P., Peletier, R.~F., Jimenez-Vicente, J., {et~al.} 2006,
  Monthly Notices of the Royal Astronomical Society, 371, 703,
  \dodoi{10.1111/j.1365-2966.2006.10699.x}

\bibitem[{Saracco {et~al.}(2023)Saracco, Barbera, De Propris, Bevacqua,
  Marchesini, De Lucia, Fontanot, Hirschmann, Nonino, Pasquali, Spiniello, \&
  Tortora}]{saracco_star_2023}
Saracco, P., Barbera, F.~L., De Propris, R., {et~al.} 2023, Monthly Notices of
  the Royal Astronomical Society, 520, 3027, \dodoi{10.1093/mnras/stad241}

\bibitem[{Schreiber {et~al.}(2018)Schreiber, Glazebrook, Nanayakkara, Kacprzak,
  Labbé, Oesch, Yuan, Tran, Papovich, Spitler, \&
  Straatman}]{schreiber_near_2018}
Schreiber, C., Glazebrook, K., Nanayakkara, T., {et~al.} 2018, Astronomy \&
  Astrophysics, 618, A85, \dodoi{10.1051/0004-6361/201833070}

\bibitem[{Serra \& Trager(2007)}]{serra_interpretation_2007}
Serra, P., \& Trager, S.~C. 2007, Monthly Notices of the Royal Astronomical
  Society, 374, 769, \dodoi{10.1111/j.1365-2966.2006.11188.x}

\bibitem[{Spitoni {et~al.}(2017)Spitoni, Vincenzo, \&
  Matteucci}]{spitoni_new_2017}
Spitoni, E., Vincenzo, F., \& Matteucci, F. 2017, Astronomy \& Astrophysics,
  599, A6, \dodoi{10.1051/0004-6361/201629745}

\bibitem[{Suess {et~al.}(2019{\natexlab{a}})Suess, Kriek, Price, \&
  Barro}]{suess_half-mass_2019}
Suess, K.~A., Kriek, M., Price, S.~H., \& Barro, G. 2019{\natexlab{a}}, The
  Astrophysical Journal, 877, 103, \dodoi{10.3847/1538-4357/ab1bda}

\bibitem[{Suess {et~al.}(2019{\natexlab{b}})Suess, Kriek, Price, \&
  Barro}]{suess_half-mass_2019-1}
---. 2019{\natexlab{b}}, The Astrophysical Journal Letters, 885, L22,
  \dodoi{10.3847/2041-8213/ab4db3}

\bibitem[{Suess {et~al.}(2020)Suess, Kriek, Price, \& Barro}]{suess_color_2020}
---. 2020, The Astrophysical Journal Letters, 899, L26,
  \dodoi{10.3847/2041-8213/abacc9}

\bibitem[{Suess {et~al.}(2023)Suess, Williams, Robertson, Ji, Johnson, Nelson,
  Alberts, Hainline, D’Eugenio, Übler, Rieke, Rieke, Bunker, Carniani,
  Charlot, Eisenstein, Maiolino, Stark, Tacchella, \&
  Willott}]{suess_minor_2023}
Suess, K.~A., Williams, C.~C., Robertson, B., {et~al.} 2023, The Astrophysical
  Journal Letters, 956, L42, \dodoi{10.3847/2041-8213/acf5e6}

\bibitem[{Taylor {et~al.}(2010)Taylor, Franx, Brinchmann, van~der Wel, \& van
  Dokkum}]{taylor_masses_2010}
Taylor, E.~N., Franx, M., Brinchmann, J., van~der Wel, A., \& van Dokkum, P.~G.
  2010, The Astrophysical Journal, 722, 1, \dodoi{10.1088/0004-637X/722/1/1}

\bibitem[{Thomas {et~al.}(2003)Thomas, Maraston, \&
  Bender}]{thomas_stellar_2003}
Thomas, D., Maraston, C., \& Bender, R. 2003, Monthly Notices of the Royal
  Astronomical Society, 339, 897, \dodoi{10.1046/j.1365-8711.2003.06248.x}

\bibitem[{Thomas {et~al.}(2005)Thomas, Maraston, \&
  Bender}]{thomas_epochs_2005}
---. 2005, The Astronomical Journal, 621, 22, \dodoi{10.1086/426932}

\bibitem[{Tomczak {et~al.}(2014)Tomczak, Quadri, Tran, Labbé, Straatman,
  Papovich, Glazebrook, Allen, Brammer, Kacprzak, Kawinwanichakij, Kelson,
  McCarthy, Mehrtens, Monson, Persson, Spitler, Tilvi, \& van
  Dokkum}]{tomczak_galaxy_2014}
Tomczak, A.~R., Quadri, R.~F., Tran, K.-V.~H., {et~al.} 2014, The Astrophysical
  Journal, 783, 85, \dodoi{10.1088/0004-637X/783/2/85}

\bibitem[{Trager {et~al.}(2000)Trager, Faber, Worthey, \&
  González}]{trager_stellar_2000}
Trager, S.~C., Faber, S.~M., Worthey, G., \& González, J.~J. 2000, The
  Astronomical Journal, 120, 165, \dodoi{10.1086/301442}

\bibitem[{Trujillo {et~al.}(2009)Trujillo, Cenarro, de~Lorenzo-Cáceres,
  Vazdekis, de~la Rosa, \& Cava}]{trujillo_superdense_2009}
Trujillo, I., Cenarro, A.~J., de~Lorenzo-Cáceres, A., {et~al.} 2009, The
  Astrophysical Journal, 692, L118, \dodoi{10.1088/0004-637X/692/2/L118}

\bibitem[{Trussler {et~al.}(2020)Trussler, Maiolino, Maraston, Peng, Thomas,
  Goddard, \& Lian}]{trussler_both_2020}
Trussler, J., Maiolino, R., Maraston, C., {et~al.} 2020, Monthly Notices of the
  Royal Astronomical Society, 491, 5406, \dodoi{10.1093/mnras/stz3286}

\bibitem[{Valentino {et~al.}(2020)Valentino, Tanaka, Davidzon, Toft,
  Gómez-Guijarro, Stockmann, Onodera, Brammer, Ceverino, Faisst, Gallazzi,
  Hayward, Ilbert, Kubo, Magdis, Selsing, Shimakawa, Sparre, Steinhardt, Yabe,
  \& Zabl}]{valentino_quiescent_2020}
Valentino, F., Tanaka, M., Davidzon, I., {et~al.} 2020, The Astrophysical
  Journal, 889, 93, \dodoi{10.3847/1538-4357/ab64dc}

\bibitem[{van~de Sande {et~al.}(2013)van~de Sande, Kriek, Franx, van Dokkum,
  Bezanson, Bouwens, Quadri, Rix, \& Skelton}]{van_de_sande_stellar_2013}
van~de Sande, J., Kriek, M., Franx, M., {et~al.} 2013, The Astrophysical
  Journal, 771, 85, \dodoi{10.1088/0004-637X/771/2/85}

\bibitem[{van~der Wel {et~al.}(2014)van~der Wel, Franx, van Dokkum, Skelton,
  Momcheva, Whitaker, Brammer, Bell, Rix, Wuyts, Ferguson, Holden, Barro,
  Koekemoer, Chang, McGrath, Haussler, Dekel, Behroozi, Fumagalli, Leja,
  Lundgren, Maseda, Nelson, Wake, Patel, Labbe, Faber, Grogin, \&
  Kocevski}]{van_der_wel_3d-hstcandels_2014}
van~der Wel, A., Franx, M., van Dokkum, P.~G., {et~al.} 2014, The Astrophysical
  Journal, 788, 28, \dodoi{10.1088/0004-637X/788/1/28}

\bibitem[{van~der Wel {et~al.}(2021)van~der Wel, Bezanson, D’Eugenio,
  Straatman, Franx, van Houdt, Maseda, Gallazzi, Wu, Pacifici, Barisic,
  Brammer, Munoz-Mateos, Vervalcke, Zibetti, Sobral, de~Graaff, Calhau,
  Kaushal, Muzzin, Bell, \& van Dokkum}]{van_der_wel_large_2021}
van~der Wel, A., Bezanson, R., D’Eugenio, F., {et~al.} 2021, The
  Astrophysical Journal Supplement Series, 256, 44,
  \dodoi{10.3847/1538-4365/ac1356}

\bibitem[{van Dokkum \& Franx(2001)}]{van_dokkum_morphological_2001}
van Dokkum, P.~G., \& Franx, M. 2001, The Astrophysical Journal, 553, 90,
  \dodoi{10.1086/320645}

\bibitem[{van Dokkum {et~al.}(2008)van Dokkum, Franx, Kriek, Holden,
  Illingworth, Magee, Bouwens, Marchesini, Quadri, Rudnick, Taylor, \&
  Toft}]{van_dokkum_confirmation_2008}
van Dokkum, P.~G., Franx, M., Kriek, M., {et~al.} 2008, The Astrophysical
  Journal, 677, L5, \dodoi{10.1086/587874}

\bibitem[{van Dokkum {et~al.}(2010)van Dokkum, Whitaker, Brammer, Franx, Kriek,
  Labbé, Marchesini, Quadri, Bezanson, Illingworth, Muzzin, Rudnick, Tal, \&
  Wake}]{van_dokkum_growth_2010}
van Dokkum, P.~G., Whitaker, K.~E., Brammer, G., {et~al.} 2010, The
  Astrophysical Journal, 709, 1018, \dodoi{10.1088/0004-637X/709/2/1018}

\bibitem[{Villaume {et~al.}(2017)Villaume, Conroy, Johnson, Rayner, Mann, \&
  van Dokkum}]{villaume_extended_2017}
Villaume, A., Conroy, C., Johnson, B., {et~al.} 2017, The Astrophysical Journal
  Supplement Series, 230, 23, \dodoi{10.3847/1538-4365/aa72ed}

\bibitem[{Williams {et~al.}(2017)Williams, Giavalisco, Bezanson, Cappelluti,
  Cassata, Liu, Lee, Tundo, \& Vanzella}]{williams_morphology_2017}
Williams, C.~C., Giavalisco, M., Bezanson, R., {et~al.} 2017, The Astrophysical
  Journal, 838, 94, \dodoi{10.3847/1538-4357/aa662f}

\bibitem[{Williams {et~al.}(2009)Williams, Quadri, Franx, Dokkum, \&
  Labbé}]{williams_detection_2009}
Williams, R.~J., Quadri, R.~F., Franx, M., Dokkum, P.~v., \& Labbé, I. 2009,
  The Astrophysical Journal, 691, 1879, \dodoi{10.1088/0004-637X/691/2/1879}

\bibitem[{Worthey \& Ottaviani(1997)}]{worthey_h_1997}
Worthey, G., \& Ottaviani, D.~L. 1997, The Astrophysical Journal Supplement
  Series, 111, 377, \dodoi{10.1086/313021}

\bibitem[{Wuyts {et~al.}(2016)Wuyts, Wisnioski, Fossati, Schreiber, Genzel,
  Davies, Mendel, Naab, Röttgers, Wilman, Wuyts, Bandara, Beifiori, Belli,
  Bender, Brammer, Burkert, Chan, Galametz, Kulkarni, Lang, Lutz, Momcheva,
  Nelson, Rosario, Saglia, Seitz, Tacconi, Tadaki, Übler, \& van
  Dokkum}]{wuyts_evolution_2016}
Wuyts, E., Wisnioski, E., Fossati, M., {et~al.} 2016, The Astrophysical
  Journal, 827, 74, \dodoi{10.3847/0004-637X/827/1/74}

\bibitem[{Zahid {et~al.}(2016)Zahid, Geller, Fabricant, \&
  Hwang}]{zahid_scaling_2016}
Zahid, H.~J., Geller, M.~J., Fabricant, D.~G., \& Hwang, H.~S. 2016, The
  Astrophysical Journal, 832, 203, \dodoi{10.3847/0004-637X/832/2/203}

\bibitem[{Zhuang {et~al.}(2023)Zhuang, Leethochawalit, Kirby, Nightingale,
  Steidel, Glazebrook, Barone, Skobe, Sweet, Nanayakkara, Allen, G.~C., Jones,
  Kacprzak, Tran, \& Jacobs}]{zhuang_glimpse_2023}
Zhuang, Z., Leethochawalit, N., Kirby, E.~N., {et~al.} 2023, The Astrophysical
  Journal, 948, 132, \dodoi{10.3847/1538-4357/acc79b}

\end{thebibliography}
\appendix
\section{Heavy metal spectra and best-fit models}
In Figures~\ref{fig:spectra_app1}$-$\ref{fig:spectra_app3} we present the LRIS+MOSFIRE spectra for all 19 galaxies from the Heavy Metal survey (black) along with their uncertainty (gray). The spectra were binned such that each pixel represents 5\;\AA in the rest frame. We show the corresponding best-fit \texttt{alf} model for each galaxy. The fitting was accomplished prior to binning the spectra. Galaxies with (*) in front of the listed ID numbers are too young for reliable abundance measurements, galaxies with ($^\wedge$) had inaccurate photometric redshifts and therefore we miss key absorption features, and galaxies with (\#) have unconstrained parameters. All (five) galaxies with symbols in front of their IDs were removed from the main analysis of this paper.

\label{app:spectra}

\begin{figure}
    \centering
    \includegraphics[width=\textwidth]{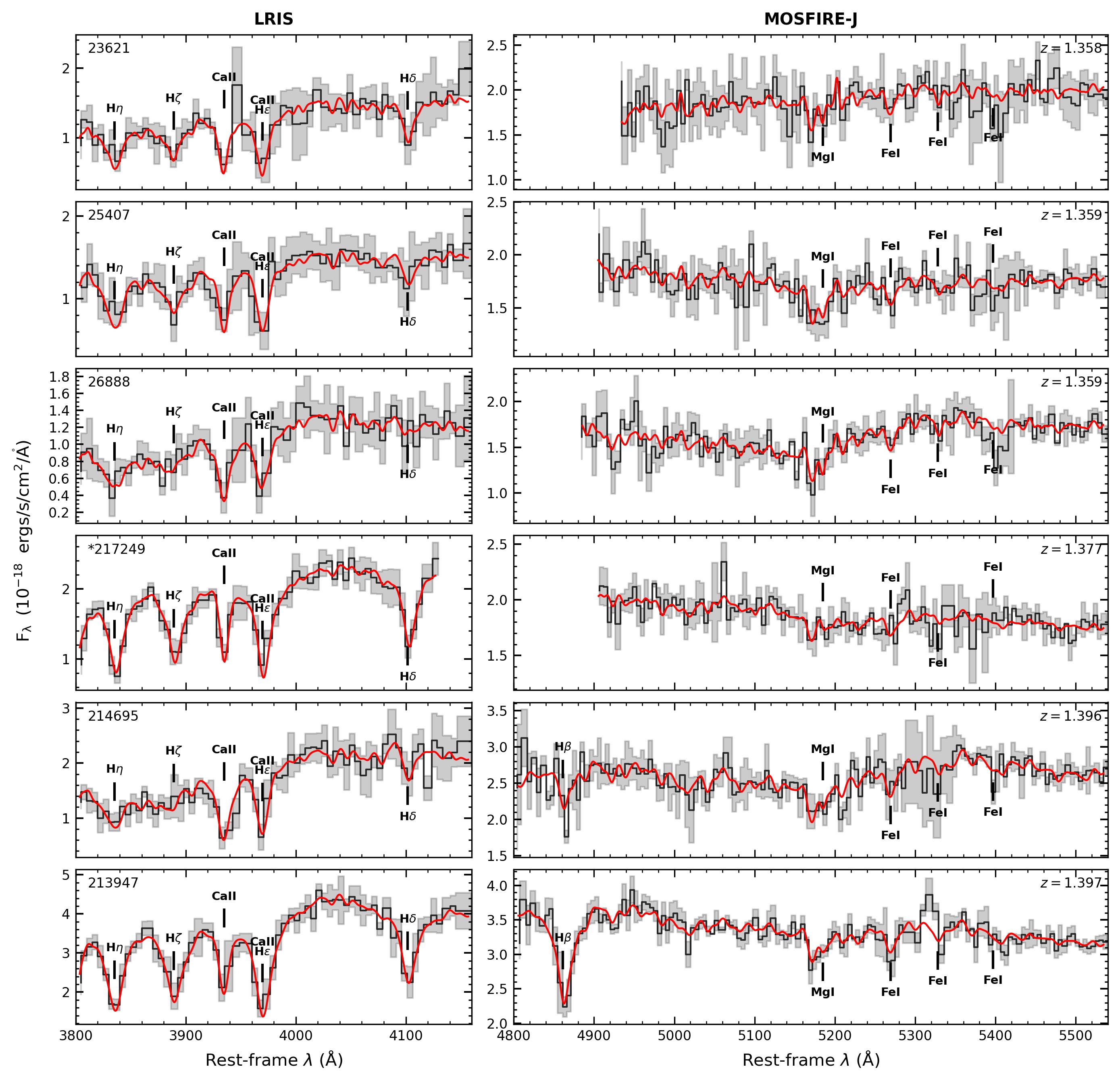}
    \caption{Spectra (black) and corresponding 1$\sigma$ uncertainties on the flux (gray), along with the best-fitting \texttt{alf} models used to derive the ages and elemental abundances (red) of primary Heavy Metal Survey targets, sorted by redshift. Spectra are binned such that one pixel is $\simeq5\,\mathrm{\mathring{A}}$. The spectra were fit prior to binning.}
    \label{fig:spectra_app1}
\end{figure}

\begin{figure}
    \ContinuedFloat
    \captionsetup{list=off,format=cont}
    \centering
    \includegraphics[width=\textwidth]{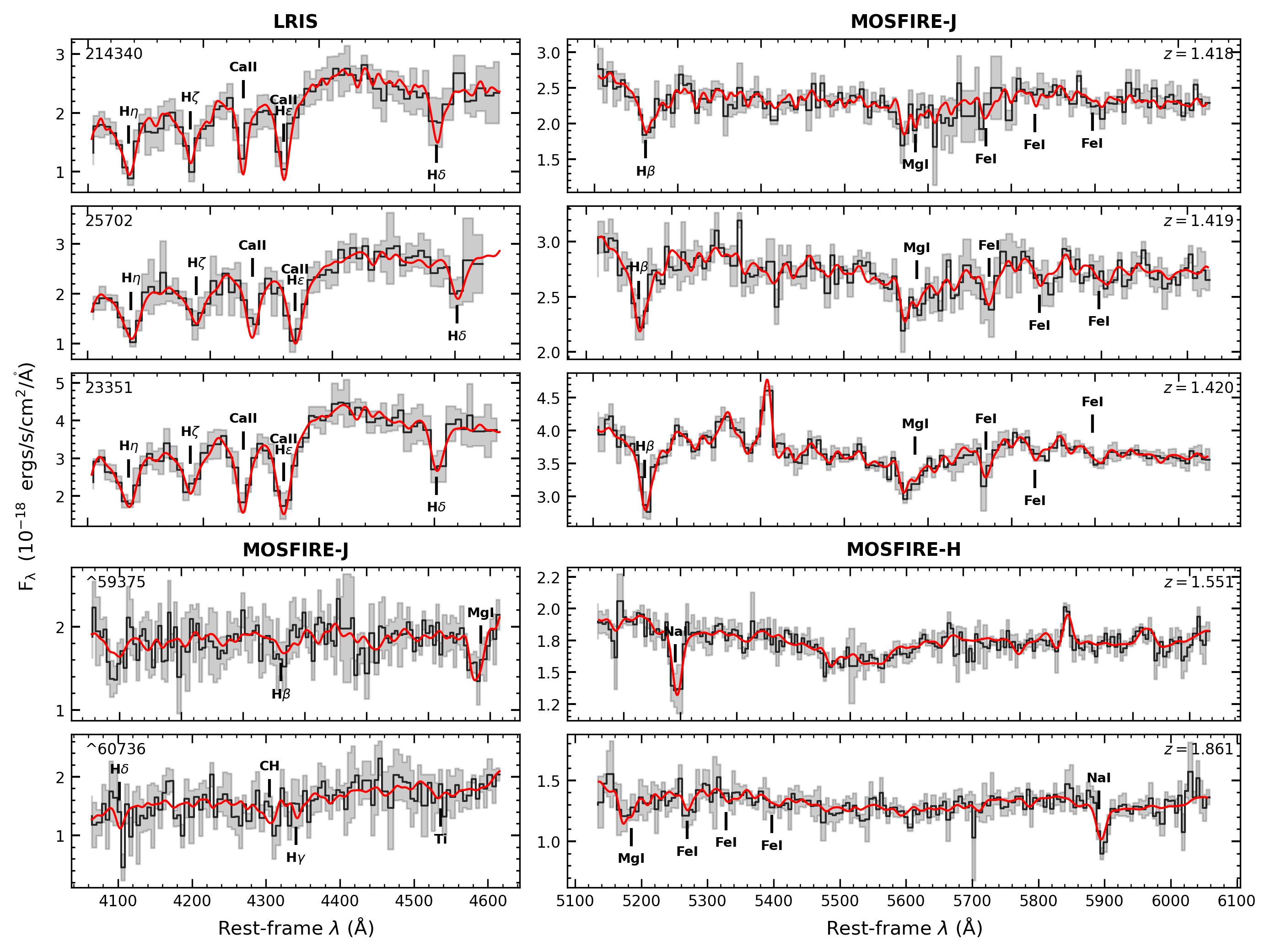}
    \caption{Continued}
    \label{fig:spectra_app2}
\end{figure}

\begin{figure}
    \ContinuedFloat
    \captionsetup{list=off,format=cont}
    \centering
    \includegraphics[width=\textwidth]{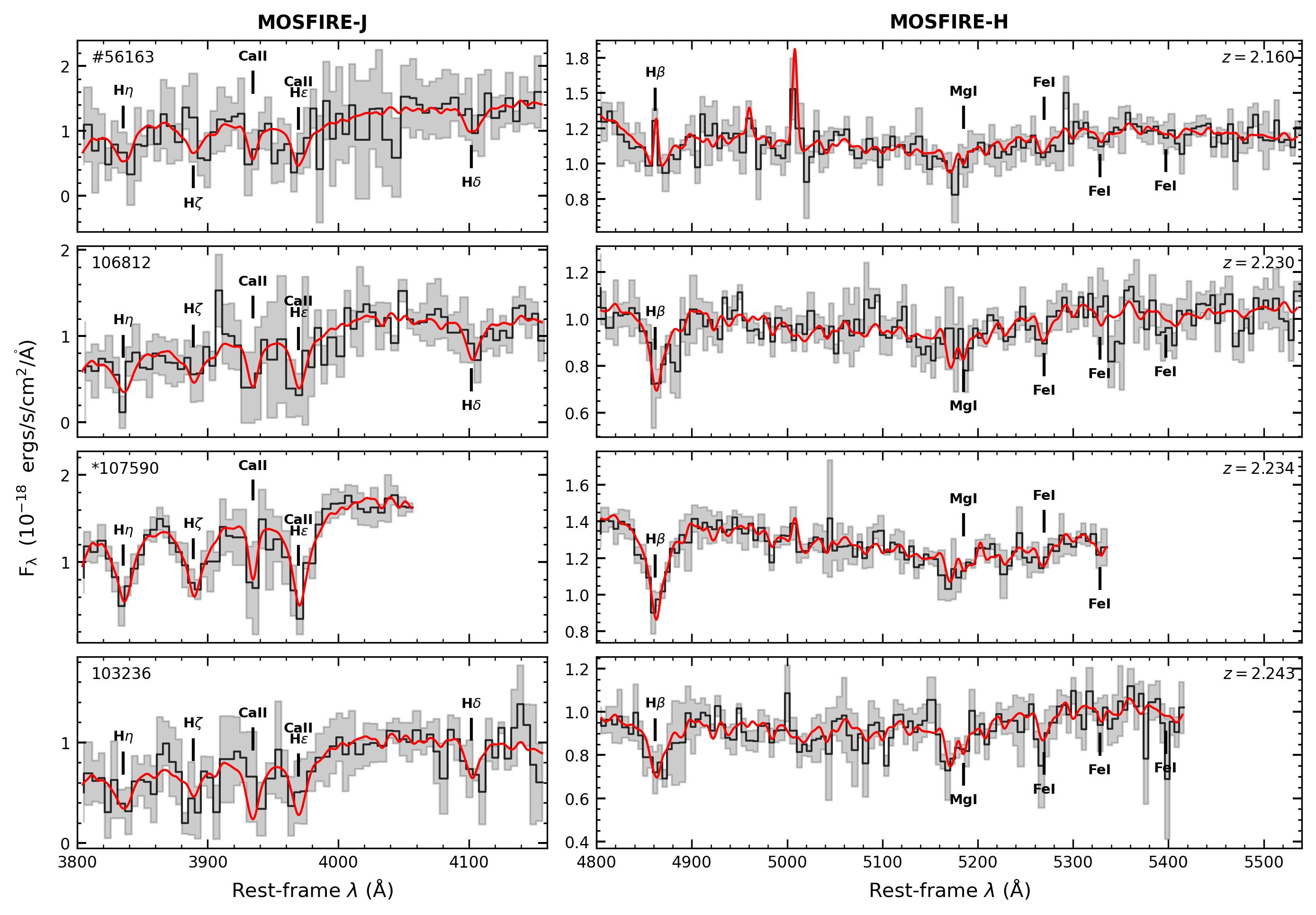}
    \caption{Continued}
    \label{fig:spectra_app3}
\end{figure}

\section{Quality of Stellar Population Modeling: Example Corner Plots}
\label{app:corner_plots}

\begin{figure}
    \centering
    \includegraphics[width=\textwidth]{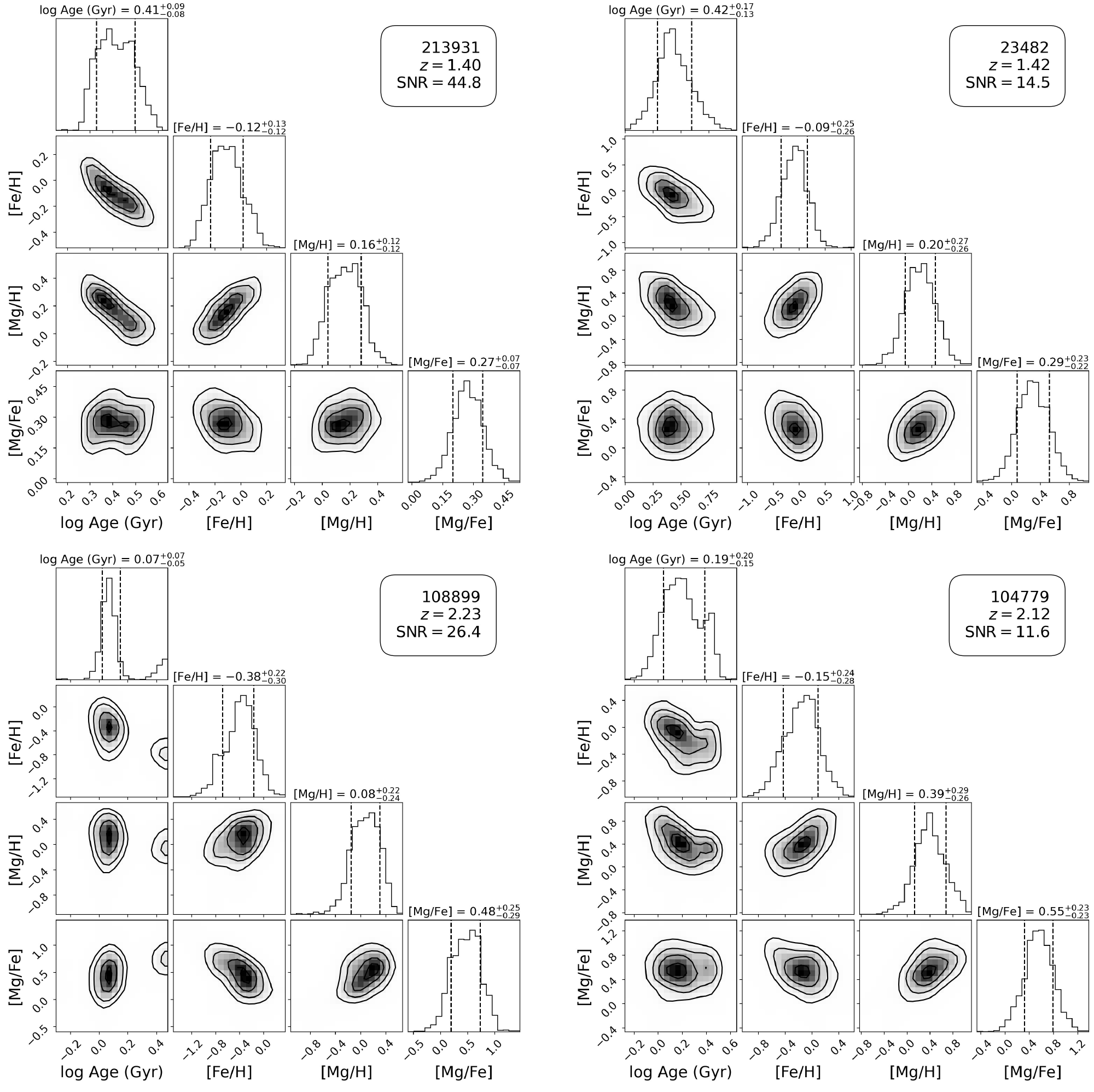}
    \caption{Four example corner plots from the full-spectrum \texttt{alf} modeling. In each panel, we show the probability distributions four different physical parameters. The dashed lines in the 1D distributions show the 1$\sigma$ levels. For each corner plot we report the galaxy ID, redshift, and SNR (\AA$^{-1}$) of the observations. }
    \label{fig:corners}
\end{figure}

In Figure~\ref{fig:corners}, we present corner plots from the full-spectrum \texttt{alf} fitting. The four corner plots correspond to the same four spectra showcased in Figure~\ref{fig:spec1}. Next to each corner plot we provide information about the object: ID, redshift, and SNR (\AA$^{-1}$). We present the four essential parameters, [Fe/H], [Mg/H], [Mg/Fe], and the age of the stellar population. The dashed lines in the plots represent the 1$\sigma$ uncertainties. Though the uncertainties are large, the results follow a normal distribution. We find there to be a degeneracy between stellar age and metallicity. Target 213931 shows the strongest degeneracy, likely because the observations are a blend of two nearby systems, resulting in the superposition of stellar populations with different ages and metallicities. For the other galaxies, the age-Z degeneracy is considerably less pronounced.

\section{Comparison with Lick Indices}
\label{app:lick}
In this Appendix we compute absorption line indices for the Heavy Metal galaxies. The Lick/IDS system, first introduced by Faber et al. (1985) and updated by Worthey et al. (1994), is the most common method for measuring elemental abundances and stellar ages of nearby quiescent galaxies. However, as we continue to push observations to higher redshifts, absorption features become fainter and more contaminated by NIR skylines. As a result, Lick indices and their derived stellar population parameters become highly uncertain. Full-spectrum modeling was, in part, developed to account for these issues. By fitting the entire stellar continuum, noise peaks can be easily down-weighted, and all absorption features for a single element can be evaluated simultaneously. 

\begin{figure}
    \centering
    \includegraphics[width=0.5\columnwidth]{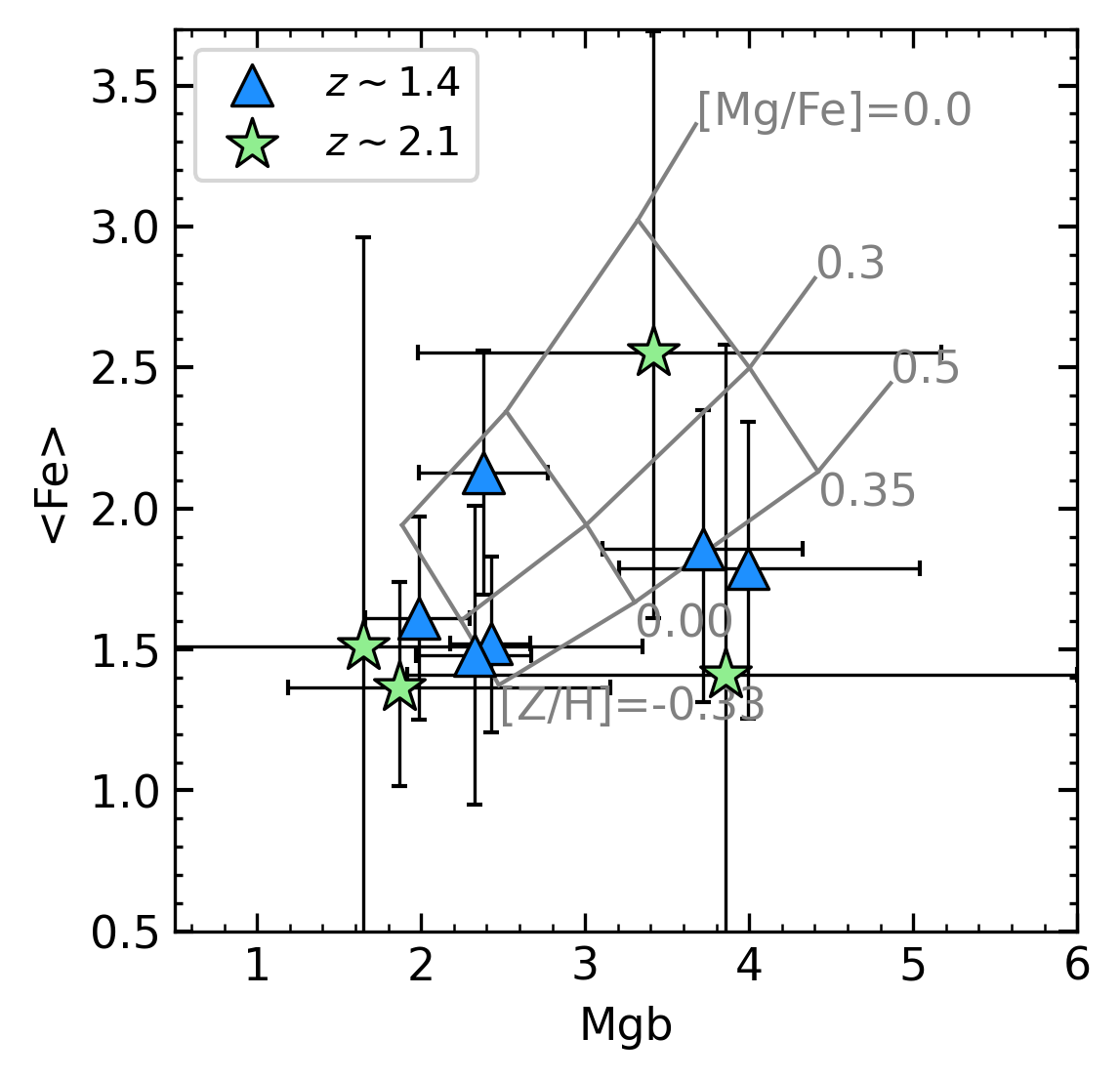}
    \caption{Lick/IDS indices Mg$b$ and $<$Fe$>$ for ten primary $z\sim1.4$ (blue triangles) and $z\sim2.1$ (green stars) Heavy Metal targets. A grid of models from \citet{thomas_stellar_2003} stellar population model ($t_{\rm age} = 2$\,Gyr) is superimposed in gray. The Heavy Metal galaxies tend towards low metallicities and high [Mg/Fe], in general agreement with what is found from the full-spectrum fitting. Uncertainties are large, motivating the need for full-spectrum modeling.}
    \label{fig:lick}
\end{figure}

In this Appendix, we check that the absorption line indices of the Heavy Metal galaxies agree with our main results from full-spectrum modeling. Specifically, we compute the $<$Fe$>$ and Mgb indices, where $<$Fe$> = \rm{0.5\times(Fe5270+Fe5335)}$. We then compare these indices to simple spectrum models from \citet{thomas_stellar_2003}. 

Given the prevalence of skylines in the Heavy Metal spectra, we must account for the prominent noise peaks. However, by default, the Lick/IDS system does not use the variance spectrum because it biases the equivalent widths. Thus, to overcome skyline contamination, we follow the approach of \citet{van_der_wel_large_2021}. For each spectrum, we take the best-fitting \texttt{alf} model and compute its $\chi^2$. We then scale the noise to a $\chi^2=1$. Any pixel that is more than $2\sigma$ or 20\% offset from the model flux is masked. If more than 30\% of the pixels in the index wavelength interval (including the pseudo-continuum intervals) are masked, the index is deemed unreliable. Four Heavy Metal galaxies are removed at this step. For the other ten galaxies, the masked pixels are replaced by a linear interpolation using the good pixels. We note that the \texttt{alf} models are \textit{only} used to help identify which pixels should be masked.

Next, we correct the absorption line indices to a constant velocity dispersion. This step is important if we want to compare indices between objects and to stellar population models. In the Lick/IDS system, indices are corrected to have zero intrinsic velocity dispersion and to have a standard wavelength-dependent resolution, ranging from 8.4\AA~to 11.5\AA~\citep{worthey_h_1997}. We compute correction factors using \texttt{alf} models at a variety of velocity dispersions \citep[e.g., see][]{davies_line-strength_1993, kuntschner_line--sight_2004}. This correction factor is applied after measuring the equivalent widths. For the Heavy Metal galaxies, the correction ranges from $C(\sigma_v)=1.0-1.5$. With the corrected Lick indices in hand, we finally estimate uncertainties by bootstrapping the spectrum and re-computing equivalent widths 10,000 times, taking the 16th and 84th percentile as the index uncertainty.

In Figure~\ref{fig:lick}, we present $<$Fe$>$ as a function of Mg$b$ for 10 Heavy Metal galaxies. We compare these indices with a set of 2 Gyr models from \citet{thomas_stellar_2003}. Drawing conclusions from this figure is challenging, as numerous Heavy Metal galaxies display index measurements with substantial uncertainties that span the entire model grid. This issue is particularly pronounced for galaxies at a redshift $z\sim2.1$. The increased uncertainties at higher redshifts can be attributed to the younger ages of these galaxies, resulting in less pronounced metal lines. Additionally, significant contamination from skylines in the near-infrared underscores the necessity for full-spectrum modeling, as opposed to index fitting, at higher redshifts.

\section{Parameter recovery with mock heavy metal spectra}
\label{app:recovery}

\begin{figure}
    \centering
    \includegraphics[width=\textwidth]{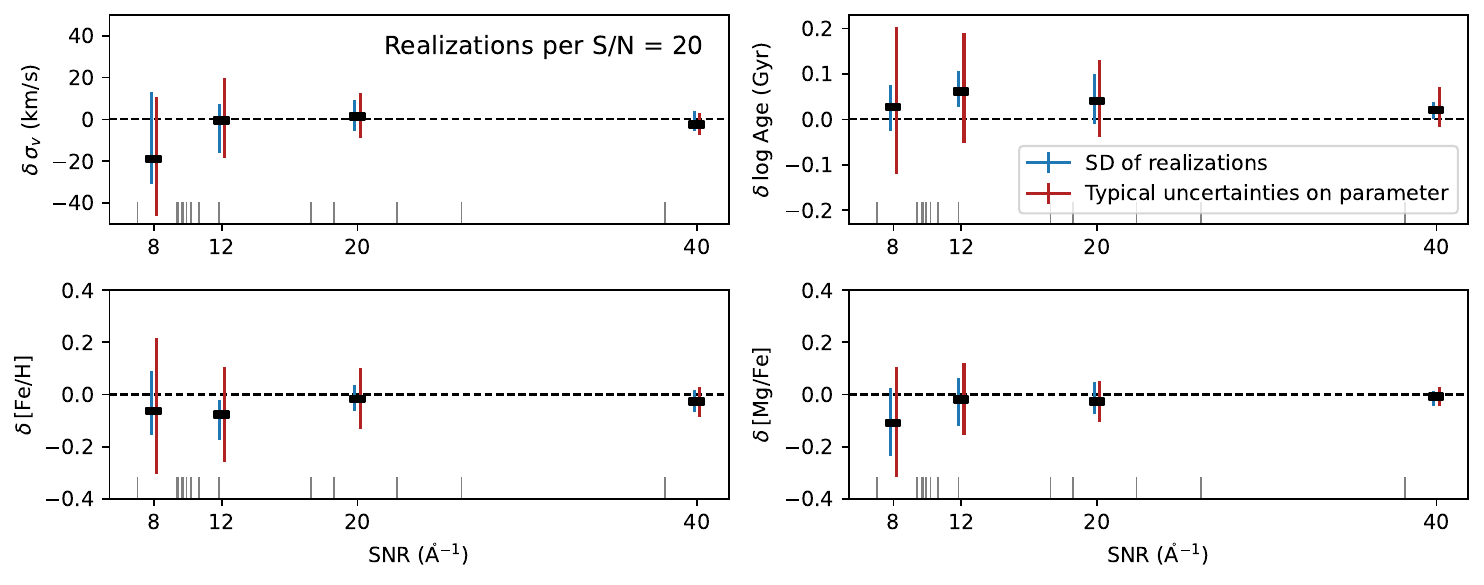}
    \caption{Results from the parameter recovery of mock Heavy Metal spectra. We show systematic offsets in the stellar population fitting code as a function of SNR for four key parameters. At each SNR, we generate 20 mock spectra and fit their elemental abundances and ages following the procedure in Section~\ref{sec:methods}. The black rectangles represent the median offset between the simulated and best-fit values for the 20 realizations. The blue errorbars represent the 16th and 84th percentiles of the best-fit values for the 20 realizations. The red errobars show the typical uncertainties on a single realization. The gray tick marks at the bottom of each panel show the true SNR of the individual Heavy Metal spectra. For the typical Heavy Metal galaxy, the systematic offsets are $\approx$-0.05\,dex for [Fe/H] and [Mg/Fe] and $\approx$0.05\,dex for $\log$\,Age. }
    \label{fig:simulation}
\end{figure}

In this Appendix, we test the robustness of our results as a function of SNR. The primary goal of this test is to check for any systematic uncertainties in full-spectrum modeling, especially as we push the limits of this fitting to lower SNRs. We begin by generating mock Heavy Metal spectra at various SNRs, fitting them using \texttt{alf}, and finally by comparing the best-fit values and their uncertainties to the true values. It is worth noting that \citet{choi_assembly_2014} conducted a similar test in their Appendix A. However, there are a few notable differences in our approach. Firstly, \texttt{alf} has undergone significant updates since the work of \citet{choi_assembly_2014}. Secondly, while \citet{choi_assembly_2014} used a flat noise spectrum for their simulated spectra, we use actual LRIS+MOSFIRE noise spectra. And finally, the Heavy Metal galaxies have much younger stellar ages, and therefore less pronounced absorption features, than what was tested in \citet{choi_assembly_2014}.

All mock spectra are generated from a single \texttt{alf} model with stellar population parameters reflecting average values from Table~\ref{table:alf-results}. We degrade the model spectrum to the desired SNR at rest-frame 5000\AA~using the Heavy Metal noise spectrum. In this way, we can replicate the observations, accounting for wavelength-dependent sensitivities. The model spectrum is also resampled to the LRIS+MOSFIRE resolution using \texttt{spectres} \citep{carnall_spectres_2017}. We generate 45 realizations at SNRs = 8, 12, 20, 40 and fit them using the same method described in Section~\ref{sec:alf}. For all realizations we compute the difference between the best-fit parameter and the known model parameters. 

We present the results in Figure~\ref{fig:simulation}. The black rectangles show the average parameter offset for the 45 realizations, with the blue errorbars representing the 1$\sigma$ scatter of these offsets. We also show the typical uncertainties on a single realization in red. The gray tick marks at the bottom of each panel represent the SNR of the individual Heavy Metal spectra. All but one Heavy Metal spectrum has SNR\;$\gtrsim10{\rm\AA^{-1}}$. $\sigma_v$ is recovered extremely well with no systematic uncertainties at SNR$\gtrsim12$. The stellar ages are also recovered well, with only a slight ($\sim0.05\;$dex) systematic offset towards older ages at SNR$\lesssim20$. The elemental abundance recovery also shows very little systematic uncertainty. There is a $-0.1\;$dex offset in [Fe/H] at SNR$\;=8$, but at the typical SNR of the Heavy Metal galaxies (SNR$>12$) there is no appreciable systematic offset to within 0.05\;dex. Similarly, [Mg/Fe] shows a slight systematic offset of 0.05\;dex at SNR$\;=8$ and SNR$\;=12$, though at larger SNR there is no appreciable systematic effect. 

This parameter recovery test confirms that the abundance results corresponding to the Heavy Metal spectra (SNR\;$\gtrsim12{\rm\AA^{-1}}$) have systematic uncertainties of $\lesssim0.05\;$dex and can be constrained to within at least 0.15\;dex. Furthermore, the results are mostly consistent with \citet{choi_assembly_2014}, despite major updates to \texttt{alf} and mock spectra generated using real LRIS+MOSFIRE noise spectra.

\end{document}